\documentclass[iop]{emulateapj}

\usepackage[colorlinks=true,linkcolor=blue,
  anchorcolor=blue,citecolor=blue,urlcolor=blue]{hyperref}  
\hypersetup{pdfauthor={Ho Seong HWANG}, 
  pdftitle={HectoMAP and Horizon Run 4}}

\slugcomment{Last updated: \today}
\shorttitle{HectoMAP and Horizon Run 4}
\shortauthors{Hwang et al.}

\usepackage{natbib}
\usepackage{subfigure}
\usepackage{url}

\begin{document}

\title{HectoMAP and Horizon Run 4: Dense Structures and Voids in the Real and Simulated Universe}

\author{Ho Seong Hwang\altaffilmark{1,2},
  Margaret J. Geller\altaffilmark{1}, 
  Changbom Park\altaffilmark{2},
  Daniel G. Fabricant\altaffilmark{1},
  Michael J. Kurtz\altaffilmark{1},
  Kenneth J. Rines\altaffilmark{3},
  Juhan Kim\altaffilmark{4},
  Antonaldo Diaferio\altaffilmark{5,6},
  H. Jabran Zahid\altaffilmark{1},  
  Perry Berlind\altaffilmark{1},
  Michael Calkins\altaffilmark{1},
  Susan Tokarz\altaffilmark{1}, and 
  Sean Moran\altaffilmark{1}}

\altaffiltext{1}{Smithsonian Astrophysical Observatory, 60 Garden Street, 
  Cambridge, MA 02138, USA; hhwang@kias.re.kr, mgeller@cfa.harvard.edu}
\altaffiltext{2}{School of Physics, Korea Institute for Advanced Study, 
  85 Hoegiro, Dongdaemun-gu, Seoul 02455, Korea}
\altaffiltext{3}{Department of Physics and Astronomy, 
  Western Washington University, Bellingham, WA 98225, USA}
\altaffiltext{4}{Center for Advanced Computation, 
  Korea Institute for Advanced Study, 85 Hoegiro, 
  Dongdaemun-gu, Seoul 02455, Korea}
\altaffiltext{5}{Dipartimento di Fisica, 
  Universit\`a degli Studi di Torino, V. Pietro Giuria 1, 
  I-10125 Torino, Italy}
\altaffiltext{6}{Istituto Nazionale di Fisica Nucleare (INFN), 
  Sezione di Torino, V. Pietro Giuria 1, I-10125 Torino, Italy}

\begin{abstract}
HectoMAP is a dense redshift survey of red galaxies
  covering  a 53 deg$^{2}$ strip of the northern sky. 
HectoMAP is 97\% complete for galaxies 
  with $r<20.5$, $(g-r)>1.0$, and $(r-i)>0.5$. 
The survey enables tests of the physical properties of large-scale structure 
  at intermediate redshift against cosmological models.
We use the Horizon Run 4,
  one of the densest and largest cosmological simulations
  based on the standard $\Lambda$ Cold Dark Matter ($\Lambda$CDM) model,
  to compare the physical properties of observed large-scale structures 
  with simulated ones in a volume-limited sample 
  covering 8$\times10^6$ $h^{-3}$ Mpc$^3$
  in the redshift range $0.22<z<0.44$. 
We apply the same criteria to the observations and simulations to
  identify over- and under-dense large-scale features of the galaxy distribution.
The richness and size distributions of 
  observed over-dense structures
  agree well with the simulated ones.
Observations and simulations
  also agree for the volume and size distributions 
  of under-dense structures, voids. 
The properties of the largest over-dense structure and 
  the largest void in HectoMAP are well within the distributions 
  for the largest structures drawn from 300 Horizon Run 4 mock surveys.  
Overall the size, richness and volume distributions of observed
  large-scale structures in the redshift range  $0.22<z<0.44$
  are remarkably consistent with predictions
  of the standard $\Lambda$CDM model. 
\end{abstract}


\section{INTRODUCTION}

Galaxy redshift surveys are a powerful tool of modern cosmology.
The large-scale 
  three-dimensional distribution of galaxies on scales ranging from 
  a few Mpc to a few hundreds Mpc
  contains important information about the initial fluctuations
  that shape structure in the universe.
The physical properties of the observed large-scale structure
  including size, richness and density distributions
  are useful probes of the physics of structure formation, and
  they contribute to the determination of cosmological parameters 
  including the matter density and dark energy
  (e.g., \citealt{teg04,eis05}).
Recent analysis of large-scale galaxy clustering
  in combination with other observations
  including the cosmic microwave background
  and Type Ia Supernovae
  suggests that we live in a universe 
  consistent with the $\Lambda$ Cold Dark Matter ($\Lambda$CDM) model,
  a homogeneous, isotropic universe 
  with Gaussian primordial fluctuations 
  dominated by dark energy and with the matter density of $\sim30\%$
  (see \citealt{wei13} for references).

The largest structures in the local universe
 revealed by galaxy redshift surveys
 (e.g., the CfA Great Wall by \citealt{gh89})
   are an additional test of the standard paradigm \citep{park90}.
As the volume covered by redshift surveys has increased, 
 larger structures were identified 
 including the Sloan Great Wall at $z\sim0.08$
 and extending for $\sim$320 $h^{-1}$ Mpc \citep{gott05}.
The existence of these large structures, 
  especially if there are even larger structures,
  might be a challenge 
 to current models of structure formation
 \citep{spr06,sd11,park12lss,park15lss}.

Cosmic voids, vast low density space in the universe,
 are also an important probe of structure formation models
 \citep{mp98,sv04, park12lss, cai15}.
Identification of voids and comparison of their properties 
  with simulations is a large-scale measure of 
  the galaxy distribution that is less sensitive than 
  dense structures to complex baryonic physics.
The statistics of the physical properties of voids
  (e.g., distributions of size, shape, or volume)
  can provide useful constraints on the cosmological parameters
  \citep{sut12,jen13}, modified gravity models \citep{cla13},
  and dark energy models \citep{bos12, pis15}.
However, identification of voids and their measured properties
  are sensitive to the survey parameters, particularly
  the sampling density \citep{sut14}.
When the sampling density of surveys is lower, 
  small voids disappear;
  the remaining voids become larger and more spherical, and
  tend to have slightly steeper radial density profiles \citep{sut14}.
These issues can have an important impact 
  on the comparison with cosmological models.
  
Several numerical simulations follow void evolution 
  in the context of hierarchical structure formation
  \citep{col05,pad05}.
However, there are few observational studies of void evolution
  as a result of the lack of dense redshift surveys
  beyond the local universe.
\citet{con05} used the Deep Extragalactic Evolutionary Probe 2 
 (DEEP2, \citealt{dav05}) redshift survey to show that
  the voids at $z\sim0.1$ are larger than
  those at $z\sim1$, approximately as expected.
\citet{mic14} have identified voids at $0.55<z<0.9$ 
 in the VIMOS Public Extragalactic Redshift Survey 
 (VIPERS; \citealt{guz14}), 
 but complex selection effects 
 prevent detailed study of size evolution of these voids
 (see also \citealt{sut14sdss} for the voids at $z=0.43-0.7$
 from the Sloan Digital Sky Survey (SDSS; \citealt{york00})).
As \citet{sut14} emphasize the importance of sampling density
 in characterizing the physical properties of voids, 
 a redshift survey that is both dense enough and extensive enough 
 to study the evolution of voids and over-dense structures
 is necessary for exploring the properties of these structures 
 at a range of cosmic epochs.

Here we use a new survey, HectoMAP, 
  to investigate the characteristics of large-scale dense structures and voids.
HectoMAP covers a 52.97 deg$^{2}$ region of the sky \citep{gel11,gh15}.
HectoMAP is a dense redshift survey of red galaxies designed
  to explore large-scale structure in the redshift range $0.2<z<0.7$. 
The ultimate limiting magnitude of the survey will be $r = 21.3$. 
We focus the  97\% complete sample with a limiting $r = 20.5$. 
For the redshift range it covers, 
  HectoMAP currently offers a unique combination of 
  high sampling density ($\sim600$ galaxies deg$^{-2}$ for the galaxies at $r<20.5$) 
  and large volume ($3.1\times10^7$ $h^{-3}$ Mpc$^3$). 
Eventual completion of the HectoMAP survey to $r = 21.3$ 
  will nearly double the sampling density on the sky to $\sim 1200$ galaxies deg$^{-2}$. 

HectoMAP complements the Baryon Oscillation Spectroscopic Survey 
  (BOSS; \citealt{daw13}),
  a similar color-selected redshift survey,
  covering a much larger volume for the galaxies at $z<0.7$ 
  ($5.4\times10^9$ $h^{-3}$ Mpc$^3$ in a 9376 deg$^2$ region).
BOSS is designed to measure the scale of baryon acoustic oscillations 
  (BAO; \citealt{eis01,daw13}).
Many studies have reported 
  impressive detections of the baryon oscillation scale based on the BOSS data
  (e.g. \citealt{kaz10, per10,pad12}).

The sampling in the BOSS survey is significantly sparser than HectoMAP.
  BOSS includes redshifts of 1.5 million luminous galaxies at $i\leq19.9$
  over 10,000 deg$^{-2}$ (i.e. $\sim150$ galaxies per square degrees).
The denser sampling of the HectoMAP survey
  allows study of the individual rich clusters, voids and 
  the surrounding large-scale structures.

Over the next few years the HectoMAP regions will be surveyed
  with Hyper Suprime-Cam (HSC, \citealt{miy12}) on Subaru. 
The resulting weak lensing map will complement the redshift survey
  in mapping the dark matter distribution.
In a pilot survey \citep{kurtz12},
  we demonstrate that HectoMAP
  matches the sensitivity of the expected Subaru weak-lensing maps.
The comparison of weak-lensing peaks with system identified
  in redshift surveys is
  an important test of the issues limiting
  applications to the measurement of cluster masses
  and to the application of weak lensing cluster counts 
  as a cosmological tool
  \citep{gel10,gel14,van13,hwa14}.
  
Here we use HectoMAP to delineate
  large-scale features in a volume-limited subsample of the survey 
  covering the redshift range $0.22 < z < 0.44$.
We compare the results with structures identified in exactly the same way from
  the Horizon Run 4 cosmological $N$-body simulations \citep{kim15sim}. 
We compare distributions of the characteristics of dense structures and voids. 
We also investigate the properties of the largest dense structure and 
  the largest void in both the real universe and the simulations.

The Horizon Run 4 is one of  the densest and largest simulations;
  there are 6300$^3$ particles in a cubic box of 
  $L_{\rm box}=3150$ $h^{-1}$ Mpc with 
  a minimum subhalo mass of 2.7$\times$10$^{11}$ $h^{-1}$ M$_\odot$
  (30 dark matter particles). 
The simulations are large enough to provide 300 independent mock surveys 
  for comparison with the data. 
The Horizon Run 4 simulations are dark matter only,
  but the comparison with galaxy distribution on large scales
  ($\gtrsim10$ Mpc)
  is insensitive to baryonic physics \citep{kaz10,vanD11,vanD14}.

We describe the HectoMAP observations in Section \ref{data}
  and the Horizon Run 4 simulations in Section \ref{nbody}.
We explain the method and the sample 
  where we identify large-scale structure
  in HectoMAP in Section \ref{identify}. 
We apply the same procedure we use for the HectoMAP data 
  to the Horizon Run 4 data in Section \ref{simiden}.
Statistical comparisons between the observations and simulations 
  are in Section \ref{results}.
We compare the largest structures in HectoMAP and Horizon Run 4 in Section \ref{big}.
We discuss the results and conclude 
  in Sections \ref{discuss} and \ref{sum}, respectively.
Throughout,
  we adopt flat $\Lambda$CDM cosmological parameters:
  $H_0 = 100$ $h$ km s$^{-1}$ Mpc$^{-1}$, 
  $\Omega_{\Lambda}=0.74$, and $\Omega_{m}=0.26$ 
  (WMAP 5-year data, \citealt{dun09}).
All quoted errors in measured quantities are 1$\sigma$.

\section{HectoMAP OBSERVATIONS}\label{data}

HectoMAP is a redshift survey of red galaxies,
  covering a 52.97 deg$^{2}$ strip of the sky with
  200$\arcdeg \leq$ R.A.(J2000) $\leq$ 250$\arcdeg$ and 
  42.5$\arcdeg \leq$ Decl.(J2000) $\leq$ 44.0$\arcdeg$.  
\citet{gel11} and \citet{gh15} preview the survey
  and describe its initial goals.
The ultimate goal of the survey is completion to $r = 21.3$. 
Here we analyze the complete 
 brighter portion of the survey for galaxies with $r<20.5$. 

\subsection{Photometry}

We used SDSS DR7 photometry (DR7, \citealt{aba09}) 
  to select targets for spectroscopic observations in 2010. 
Since 2013 we have used the updated DR9 photometric catalog 
  \citep{ahn12}\footnote{The photometric data
  in SDSS DR12 are the same as in DR9.}.
Targets are red galaxies with $r_{\rm Petro,0}<21.3$,
  $(g-r)_{\rm fiber,0}>1.0$, $(r-i)_{\rm fiber,0}>0.5$, and
  $r_{\rm fiber,0}<22.0$\footnote{The subscript, ``0'', denotes
  Galactic extinction corrected magnitudes. 
  The ``fiber'' and ``Petro'' indicate SDSS 
  fiber and Petrosian magnitudes, respectively.}.
We examined a somewhat broader color selection in a pilot survey
  ($g-r>1.0$, $r-i>0.4$ and $r_{\rm Petro}<21.3$; \citealt{kurtz12})
  where we compared apparent clusters in the redshift survey 
  with weak-lensing peaks.
Red galaxies satisfying the color selection 
  we used are certainly a robust basis for cluster identification 
 (see also \citealt{gh15} for detailed discussion 
  on the color selection criteria). 
In general, red galaxies are more clustered and the contrast
  of the large-scale structure is greater than blue galaxies 
  (e.g., \citealt{mad03,park07,zeh11}).
The color selection efficiently rejects nearby galaxies 
  at $z\le0.2$ where the overlap with the SDSS is substantial and 
  where the lensing sensitivity drops steeply 
  (see \citealt{gel10}).

\begin{deluxetable}{lccr}
\tablewidth{0.48\textwidth} 
\tablecaption{\large HectoMAP Redshift Survey Properties 
\label{table1}}
\tablehead{
Parameter & Value }
\startdata
   Survey Area (deg$^2$) & 52.97\\
   N$_{\rm phot, 20.5, mag}$\tablenotemark{a} & 32808 \\
   N$_{\rm z, 20.5, mag}$\tablenotemark{b}& 31721 \\ 
   $z_{\rm med, 20.5, mag}$\tablenotemark{c}&0.34\\
   N$_{\rm z, 20.5, vol}$\tablenotemark{d}& 9881 \\ 
\enddata
\tablenotetext{a}{Number of photometric objects 
  in the bright sample of red galaxies with 
  $(g-r)_{\rm fiber,0}>1.0$, $(r-i)_{\rm fiber,0}>0.5$, 
  $r_{\rm fiber,0}<22.0$ and  $r_{\rm Petro,0}<20.5$.}
\tablenotetext{b}{Number of galaxies with a measured redshift.} 
\tablenotetext{c}{Median redshift of the galaxies.}
\tablenotetext{d}{Number of galaxies in the volume-limited sample 
  (see Section \ref{volsamp}).}
\end{deluxetable}

\subsection{Spectroscopy}\label{survey}

We obtained galaxy spectra with the Hectospec 
  on the MMT 6.5m telescope \citep{fab98,fab05}.
The Hectospec is a robotic instrument with 300 fibers 
  deployable over 1$\arcdeg$ field of view.
We used the 270 line mm$^{-1}$ grating of Hectospec that
  provides a dispersion of 1.2 $\AA$ pixel$^{-1}$ and 
  a resolution of $\sim$6 $\AA$.
Typical exposures were 0.75$-$1.5 hours. The resulting spectra
  cover the wavelength range 3650$-$9150 $\AA$.

We chose the HectoMAP strip at high declination
  (i.e. 42.5$\arcdeg \leq$ Decl. $\leq$ 44.0$\arcdeg$)
  so that it is always 30$\arcdeg$ away from the moon.
Thus we can observe  target galaxies 
  in gray  or even in bright time. 
The advantage of this high declination location is particularly favorable 
  for the brighter portion of the survey we consider here.

To obtain a high, uniform spectroscopic completeness
  within the survey region
  down to the ultimate limiting magnitude of $r_{\rm Petro,0}=21.3$,
  we weighted the spectroscopic targets 
  with their apparent magnitudes within the color range.
We used the Hectospec observation planning software \citep{roll98}
  to assign the fibers efficiently.
We filled unused fibers with targets bluer than the color limits.
  
We reduced the Hectospec spectra obtained before 2013
  with the \citet{mink07} pipeline. 
Beginning in 2014, we used HSRED v2.0,
  an updated reduction pipeline 
  originally developed by Richard Cool.
There is no systematic offset between
 redshifts derived from the two pipelines.
We determined the redshifts by using RVSAO \citep{km98} 
  to cross-correlate the spectra with templates.
We visually inspected all of the spectra, and 
  assigned a quality flag to the spectral fits with
  ``Q'' for high-quality redshifts, ``?'' for marginal cases, and
  ``X'' for poor fits.
We use only the spectra with ``Q'' redshifts.
Repeat observations of 1651 separate absorption-line and 
  238 separate emission-line
  objects provide mean internal errors normalized by $(1+z)$
  of 48 and 24 km s$^{-1}$, respectively \citep{gel14f2}.

\begin{figure}
\center
\includegraphics[width=0.48\textwidth]{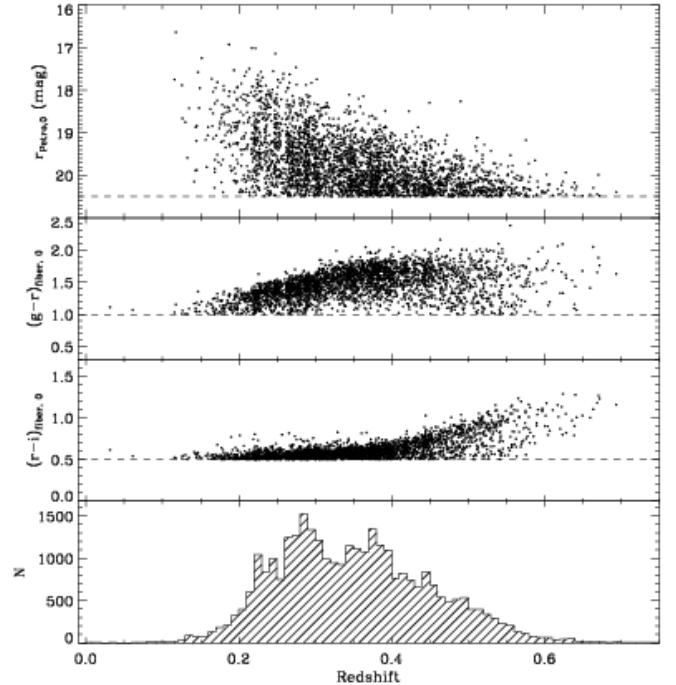}
\caption{({\it Top}) $r$-band magnitudes,
 ({\it Middle}) ($g-r$) and ($r-i$) colors as a function of redshift
 for HectoMAP galaxies with
 $(g-r)_{\rm fiber,0}>1.0$, $(r-i)_{\rm fiber,0}>0.5$,
 $r_{\rm fiber,0}<22.0$, and  $r_{\rm Petro,0}<20.5$.
The horizontal line in each panel indicates these selection criteria.
All magnitudes are Galactic-extinction corrected.
We display only 10\% of the data for clarity.
({\it Bottom}) Redshift distribution for HectoMAP. 
}\label{fig-color}
\end{figure}

\begin{figure}
\center
\includegraphics[width=0.48\textwidth]{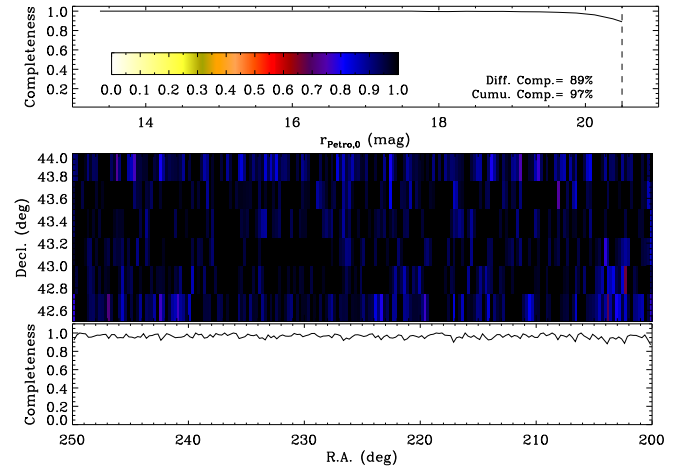}
\caption{Spectroscopic completeness for the sample of
   galaxies with $r_{\rm Petro,0}<20.5$
  as a function of $r$-band magnitude ({\it Top}) and 
  of right ascension ({\it Bottom}).
Vertical line in the top panel is the magnitude limit, $r_{\rm Petro,0}=20.5$.
({\it Middle}) Two-dimensional spectroscopic completeness
  as a function of right ascension and of declination.  
}\label{fig-comp}
\end{figure}

\begin{figure*}
\center
\includegraphics[width=0.99\textwidth]{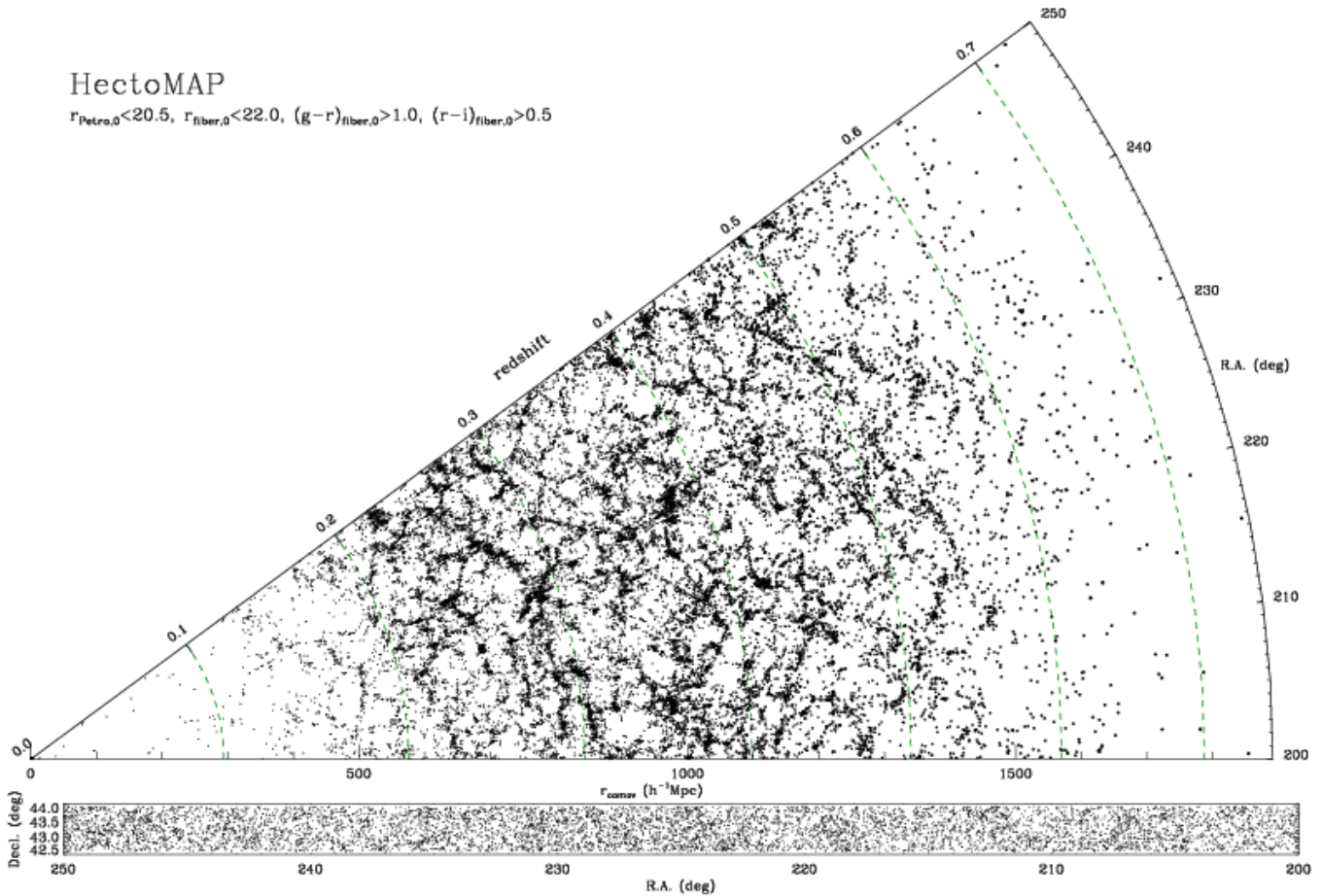}
\caption{({\it Top}) Cone diagram for HectoMAP galaxies with  $r_{\rm Petro,0}<20.5$.
({\it Bottom}) Distribution on the sky  
  as a function of right ascension and  declination.
We display all the data in the top panel, but
  only 30\% of the data for clarity in the bottom panel.
}\label{fig-cone}
\end{figure*}

We observed the HectoMAP region in queue mode beginning in 2010. 
We expect to complete the survey 
  to the deeper limiting magnitude $r = 21.3$ in Spring 2016.
We supplement the data with redshifts from
  the SDSS DR12 \citep{sdssdr12}
  and the NASA/IPAC Extragalactic Database (NED) including 
  the literature \citep{gro04,jaf13}.
HectoMAP is now 97\% complete to $r = 20.5$ and includes 31721 redshifts.
Among them, 30555 were acquired with Hectospec,
   1165 were measured by the SDSS DR12 \citep{sdssdr12},
   and one was from \citet{gro04}.
Table \ref{table1} lists the number of galaxies 
  in the photometric sample with $r_{\rm Petro,0}<20.5$
  and the number of measured redshifts in the sample. 

Figure \ref{fig-color} shows the HectoMAP color selection.
Beginning from the top panel, 
  we display, sequentially,  the distributions of $r$-band apparent magnitudes,
  $(g-r)_{\rm fiber,0}$ and $(r-i)_{\rm fiber,0}$ colors.
We plot only the galaxies satisfying the color selection.
The histogram clearly demonstrates that the color selection 
  efficiently rejects nearby galaxies with $z<0.2$.
The median redshift of the red galaxy sample is $z\sim0.34$.

The upper panel of Figure \ref{fig-comp} shows 
 the spectroscopic completeness
 for the sample with 
  $(g-r)_{\rm fiber,0}>1.0$, $(r-i)_{\rm fiber,0}>0.5$, 
  $r_{\rm fiber,0}<22.0$ and  $r_{\rm Petro,0}<20.5$
  as a function of apparent magnitude.
The completeness curve is nearly flat and decreases only slightly 
  at $r_{\rm Petro,0}>20$.
The cumulative spectroscopic completeness for 
  $r_{\rm Petro,0}<20.5$
  is 97\% with a differential completeness 
  of 89\% at $r_{\rm Petro,0}=20.5$.
The middle panel shows a two-dimensional map of the completeness  
  at $r_{\rm Petro,0}<20.5$ as a function of R.A. and decl.
The two-dimensional map shows 200$\times$6 pixels
  for the survey region of 
  200$\arcdeg \leq$ R.A.(J2000) $\leq$ 250$\arcdeg$ and 
  42.5$\arcdeg \leq$ Decl.(J2000) $\leq$ 44.0$\arcdeg$.
The map shows that the survey is highly complete
  over the survey region;
  there are only 37 pixels (3.1\%) with completeness less than 85\%. 
The bottom panel shows the integrated completeness 
  as a function of R.A.

\begin{figure}
\center
\includegraphics[width=0.48\textwidth]{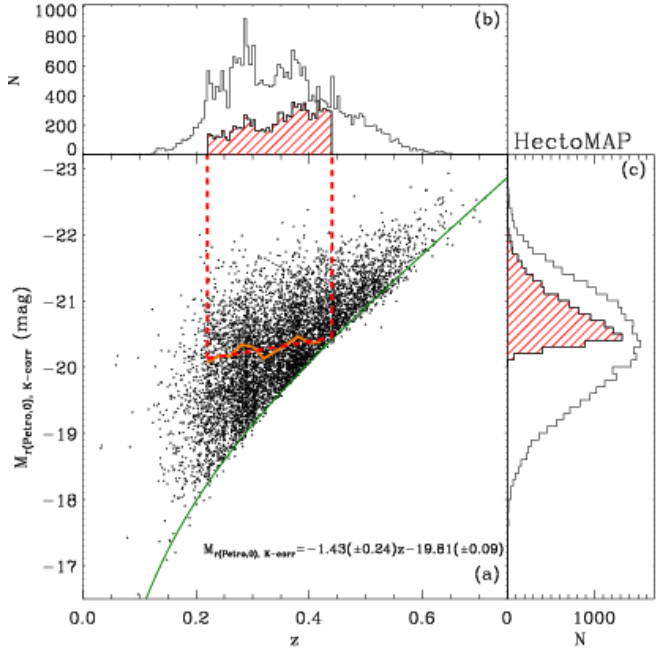}
\caption{$K$-corrected (to $z=0.4$) $r$-band absolute magnitudes
  for galaxies in HectoMAP 
  as a function of redshift.
The green curve indicates the apparent magnitude limit, 
  $r_{\rm Petro,0}=20.5$.
The orange solid line defines the lower limit for the galaxy sample that 
  has a constant galaxy number density over the redshift range 
  $0.22<z<0.44$ 
  ($n_{\rm gal}=1.36\times10^{-3}$ $h^{3}$ Mpc$^{-3}$ or 
  $d_{\rm mean}=9.01$ $h^{-1}$ Mpc).
The slanted red dashed line is the best fit to the orange line
  (see the equation in the panel); it 
  defines a volume-limited sample of galaxies 
  delimited by the vertical dashed lines.
We display only 20\% of the data for clarity.
The open and red hatched histograms in top and right panels
  show the distributions of galaxies 
  in the entire and volume-limited samples, respectively.
}\label{fig-absz}
\end{figure}

Figure \ref{fig-cone} shows a cone diagram 
  for the  sample with $r_{\rm Petro,0}<20.5$
  projected along the declination direction.
The diagram shows the characteristic features of large-scale structure 
 at intermediate redshift;
 fingers corresponding to clusters are obvious at $z<0.55$, 
 and there are many voids delimited by thin walls and filaments.
At $z>0.45$, the structures are less sharply defined
  than those at lower redshift
  because we have only very luminous galaxies 
  at this redshift range  (see Figure \ref{fig-absz}) and
  because the physical size of the survey slice is thicker 
  at higher redshift than for lower redshift; the
  1.5 degree slice corresponds to 25.7 and 12.6 $h^{-1}$ Mpc 
  at $z=0.6$ and $z=0.2$, respectively.

\section{HORIZON RUN 4 $N$-BODY SIMULATION}\label{nbody}

We use the Horizon Run 4 $N$-body simulation \citep{kim15sim} 
  as a foundation for comparing the
  large-scale features of HectoMAP with the predictions
  of structure formation models.
This dark matter only simulation is one of 
  the densest and largest available. 
It is large enough to contain many independent mock HectoMAP surveys
  and is thus ideal for the comparison.

The Horizon Run is a series of cosmological
  $N$-body simulation run by \citet{kim11sim}.
Horizon Run 4 is the densest of these simulations
  with 6300$^3$ particles in a cubic box of 
  $L_{\rm box}=3150$ $h^{-1}$ Mpc.
The simulation adopts a standard $\Lambda$CDM
  cosmology in accord with the WMAP 5-year data \citep{dun09}.
The particle mass is $m_p\sim9\times10^{9}$ $h^{-1}$ M$_\odot$.
The subhalos are identified with the physically self-bound (PSB)
  subhalo finding method \citep{kp06}.
The minimum mass of subhalos
  with 30 member particles is
  2.7$\times$10$^{11}$ $h^{-1}$ M$_\odot$.
  
Horizon Run 4 provides
  past lightcone data\footnote{The simulation
  outputs including snapshot data, 
  past lightcone data, and halo merger data are available
  at \url{http://sdss.kias.re.kr/astro/Horizon-Run4/}} up to $z\sim1.5$.
The true lightcones are important for comparison
  with a deep redshift survey like HectoMAP.
Using simulated light cone data in redshift space,
  we first construct 300 non-overlapping mock surveys. 
Each mock survey has the same survey geometry as HectoMAP.

When we compare HectoMAP with Horizon Run 4,
   we assume that each dark matter subhalo contains only one galaxy.
This subhalo-galaxy one-to-one correspondence assumption
  has worked successfully in many applications including 
  the one-point function and its local density distribution \citep{kim08}, 
  the two-point function \citep{kim09,nuza13}, 
  the topology of galaxy distribution \citep{gott09,choi10top,par14},
  estimates of large-scale velocity moments \citep{aga12}, 
  and identification of the largest structures \citep{park12lss}. 
We also match the halo number density of the mock data
  with the galaxy number density in HectoMAP  
  (i.e. abundance matching, \citealt{kra04,vo04,con06,kim08,guo10}).

To test the sensitivity of structure identification 
  to the method of matching the simulated halos to the HectoMAP galaxies,  
  we construct two types of volume-limited samples of halos 
  based on two different sampling methods.
In the first method, we match the halo number density
  with the galaxy number density by assuming that 
  more massive halos correspond to more luminous galaxies
  \citep{kra04,vo04,con06,kim08,guo10}. 
A second approach mimics the known observational selection effects 
  that exist in a red-selected sample like HectoMAP. 
Red galaxies are generally in denser regions 
 (e.g., \citealt{bla05cdr,coo06,park07}),
 and the more massive they are, 
 the denser the surroundings (e.g., \citealt{lk99,park07,haas12}).
Obviously we do not have color information for halos in the simulation.
Thus, we use the local density around red galaxies as a proxy for color. 
We measure the local density around galaxies in HectoMAP and 
  match it to the local density distribution
  around halos in the simulation.

\section{ANALYSIS OF HectoMAP }\label{identify}

We construct a volume-limited sample of HectoMAP galaxies in Section \ref{volsamp}
  for robust comparison with the $N$-body simulations.
We then identify over-dense large-scale features in Section \ref{overden}
  and under-dense structures in Section \ref{underden}.
  
\subsection{A Volume-Limited Sample of HectoMAP}\label{volsamp}

To identify large-scale structures in HectoMAP 
  for robust comparison with the $N$-body simulations,
  we first construct a volume-limited sample of galaxies
  with constant comoving number density.
This approach is similar to the construction of the sample of luminous red galaxies
  for the study of large-scale structure in SDSS/BOSS 
  (e.g., \citealt{eis01,daw13}).
This procedure is the foundation  
  for a fair comparison with the simulations based on 
  the same comoving number density of halos 
  (i.e. $1.36\times10^{-3}$ $h^3$ Mpc$^{-3}$ or mean halo separation of 9.01 $h^{-1}$ Mpc),
  and for unbiased identification of large-scale structures
  within the redshift range of the sample.
  
To construct a volume-limited sample,
  we first plot the $r$-band absolute magnitudes
  for the $r_{\rm Petro,0}<20.5$ sample
  as a function of redshift (Figure \ref{fig-absz}).
We use the Kcorrect software (ver. 4.2) of 
  \citet{bla07kcorr} for $K$-corrections.
We then compute galaxy number densities
  by changing the lower limit in absolute magnitude 
  as a function of redshift. 
The orange contour indicates the magnitude limit
  that yields a comoving number density of 
  $1.36\times10^{-3}$ $h^3$ Mpc$^{-3}$ at each redshift.
To remove the effect of small-scale fluctuations, 
  we fit the contour
  with a linear relation (slanted red dashed line).
The vertical red dashed lines show the lower and upper redshift limits
 defining the volume-limited sample we analyze.
The upper redshift limit of $z=0.44$ is set by the
  $r_{\rm Petro,0}=20.5$ (green solid line) magnitude limit 
  of the redshift survey;
  the lower redshift limit is set by the ($r-i$) selection 
  that eliminates low redshift objects.

\begin{figure*}
\center
\includegraphics[width=0.9\textwidth]{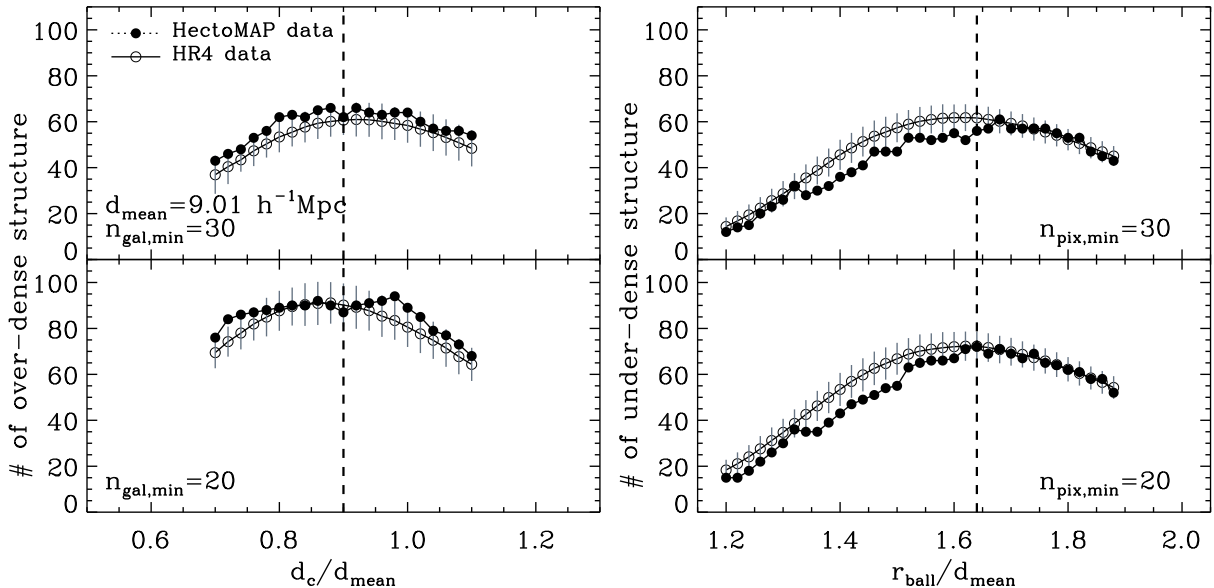}
\caption{({\it Top left}) Number of over-dense structures
  with more than 30 member galaxies
  as a function of linking length in units of galaxy mean separation
  for HectoMAP (filled circles)
  and Horizon Run 4 (open circles).
The error bars indicate the standard deviation
  in the number of structures derived from 
  the 300 mock samples of Horizon Run 4.
The vertical lines indicate the critical linking length we adopt 
  (i.e. $d_c=0.9d_{\rm mean}$).
({\it Bottom left}) Same as top left panel, but for 
  over-dense structures
  with more than 20 members.
({\it Top right}) Number of under-dense structures
  with more than 30 connected pixels
  as a function of the ball radius for HectoMAP (filled circles)
  and Horizon Run 4  (open circles).
The vertical lines indicate the critical ball radius we adopt 
  (i.e. $r_{\rm ball}=1.64d_{\rm mean}$).
({\it Bottom right}) Same as top right panel, but for 
  under-dense structures
  with more than 20 connected pixels.
}\label{fig-num}
\end{figure*}

\subsection{Identification of Over-dense Large-Scale Structures}\label{overden}

We identify over- and under-dense large-scale structures
  using the method in \citet{park12lss}.
This method is similar to the application of 
  a friends-of-friends algorithm \citep{am98,ber06}
  for identifying galaxy groups/clusters and cosmic voids.
We explain the details in this section.

We apply a friends-of-friends algorithm \citep{hg82}
  to the volume-limited sample of HectoMAP galaxies
  to identify over-dense large-scale structures 
  by connecting galaxies with a fixed linking length.
To reduce the fingers 
 along the line of sight resulting 
 from peculiar motions in galaxy systems,
 we adopt a method similar to the ones used
 for SDSS galaxies in \citet{teg04} and \citet{park12lss}.
We first run the friends-of-friends algorithm
 with a short linking length of 3 $h^{-1}$ Mpc 
 comparable with the diameter of a rich cluster.
We then compare the dispersion of 
  the linked structures along the line of sight
  with the dispersion in projected separation perpendicular to the line of sight.
If the dispersion along the line
  of sight exceeds the perpendicular spread,
  we adjust the radial velocities of the member galaxies
  to have the same effective velocity dispersion in the two directions.

We again apply the friends-of-friends algorithm
  to the data after the correction for extension 
  along the line-of-sight.
This time we explore several linking lengths 
  to identify the optimal linking length that 
  gives the maximum number of structures \citep{bas03}.
The left panels of Figure \ref{fig-num}
  show the number of over-dense structures identified
  as a function of linking length. 
The upper and lower panels show the numbers of structures
  containing more than 30 and 20 member galaxies, respectively
  (see filled circles for the HectoMAP data).
The number of over-dense structures first 
  increases with linking length and then decreases 
  for a large enough linking length.   
We choose the length $d_c=0.9d_{\rm mean}=8.11$ $h^{-1}$ Mpc that
  gives the maximum number of structures 
  as the critical linking length, and 
  set 20 as the minimum number of member galaxies 
  defining an over-dense structure.
We explore different linking lengths further in Section \ref{results}. 

\begin{figure*}
\center
\begin{tabular}{cc}
\includegraphics[width=0.99\textwidth]{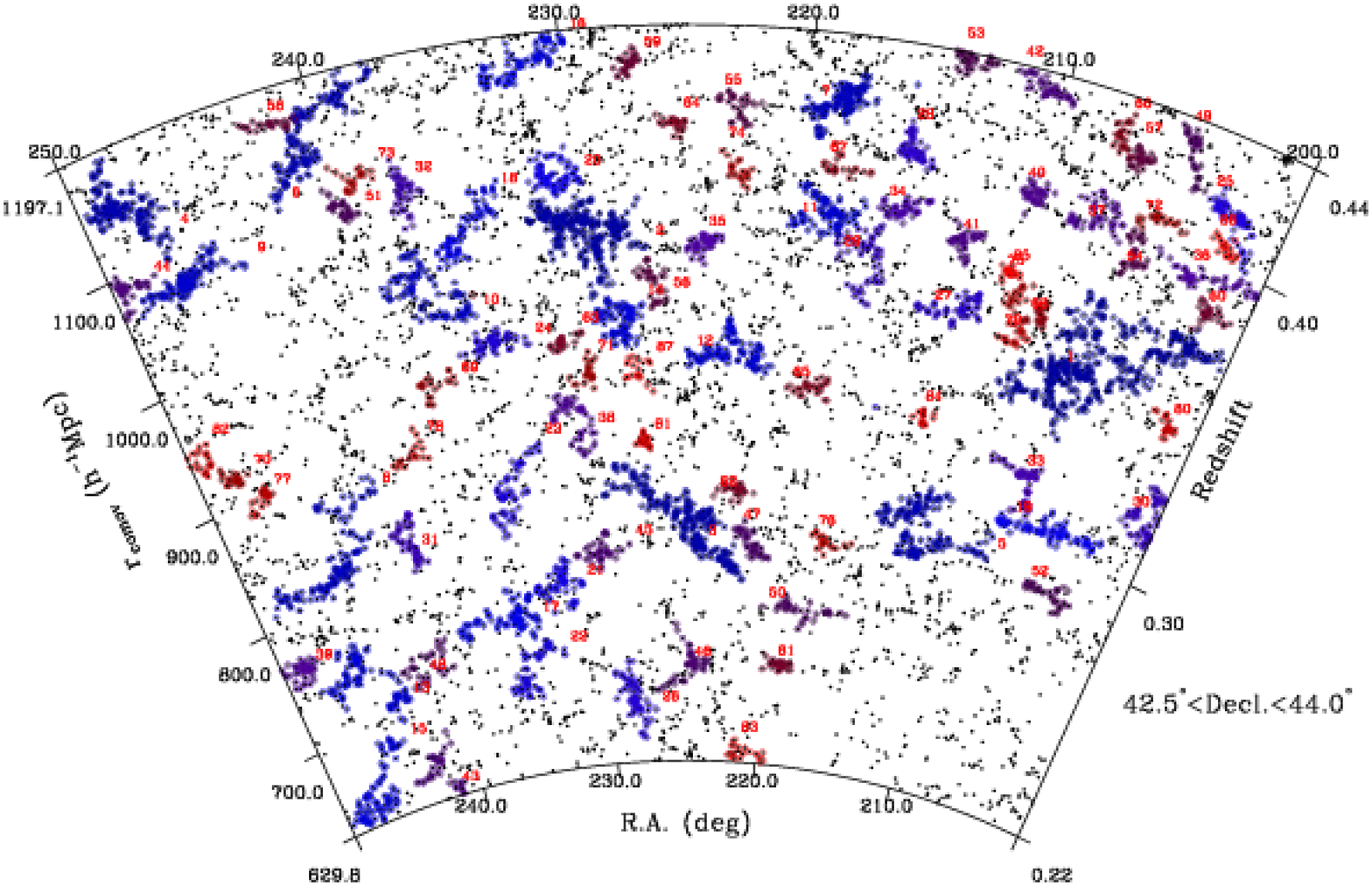} \\
\includegraphics[width=0.99\textwidth]{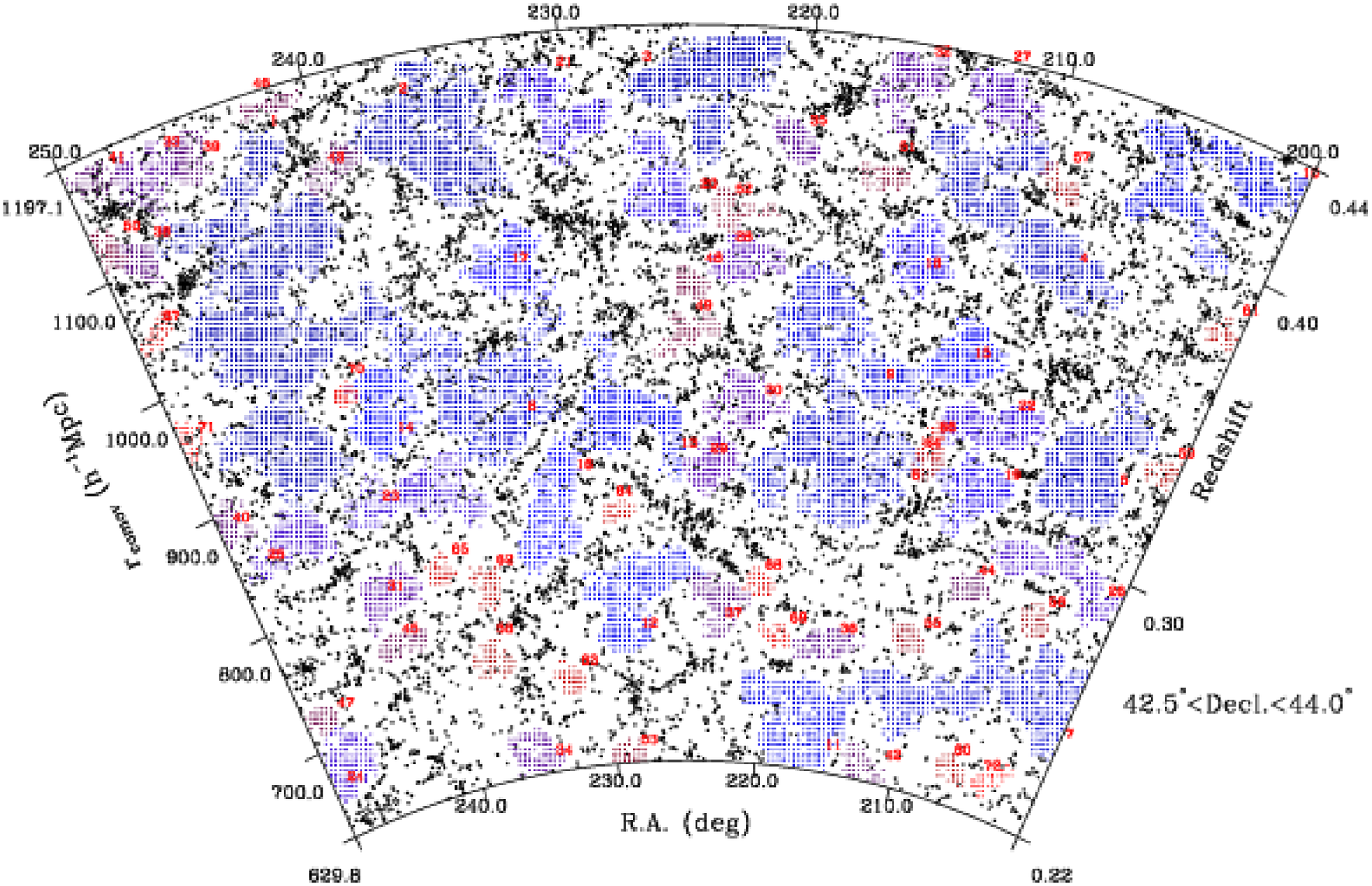}
\end{tabular}
\caption{({\it Top}) Distribution of galaxies in the volume-limited sample
  derived from HectoMAP. 
The colored circles with numbers indicate over-dense structures identified
  with the friends-of-friends algorithm.
({\it Bottom}) Same as top panel, but for under-dense structures.
}\label{fig-hpid}
\end{figure*}

The top panel of Figure \ref{fig-hpid} shows
  a cone diagram for the volume-limited sample
  of galaxies at $0.22<z<0.44$ (black dots).
We mark the members 
  of over-dense structures identified 
  with the critical linking length of $d_c=0.9d_{\rm mean}$
  with colored symbols.
Many structures we identify are filamentary.

The richest and largest structure in this diagram is 
  at R.A.$=$205.5 (deg), Decl.$=$43.2 (deg) and $z=0.36$;
  the maximum extent is 181.1 $h^{-1}$ Mpc 
  with 443 members.
The maximum extent is the maximum separation of
 member galaxies in a structure. We compute this extent in real space
 after reducing the extension of fingers in redshift space.
For comparison, \citet{park12lss} 
  derive a length of $\sim$150 $h^{-1}$ Mpc 
  for the Sloan Great Wall using the SDSS galaxies. 
However, the linking length used in Park et al. 
  is not the same as the one we use here.
They used a sample of SDSS galaxies 
  with $M_r\leq-21.6$, $z<0.17$, 
  and $d_{\rm mean}=9$ $h^{-1}$ Mpc 
  with a linking length of $0.622 d_{\rm mean}$ ($=5.6$ $h^{-1}$ Mpc).
If we use a linking length of $0.622 d_{\rm mean}$,
  the largest structure in HectoMAP
  has a maximum extent of only 60.3 $h^{-1}$ Mpc with 135 members. 
Thus the extent of the apparently largest structure 
  is a sensitive function of the linking length.
Although the difference between HectoMAP and SDSS may result partly
  from differences in survey geometry 
  (HectoMAP is essentially a 2D thin-slice survey whereas the SDSS is 3D),
  this comparison underscores the necessity of 
  using identical procedures when comparing different redshift surveys 
  or when comparing the data with a simulation.

\subsection{Identification of Voids: Under-Dense Structures}\label{underden}

We also identify under-dense large-scale features or voids 
  in HectoMAP.
We begin by tessellating the three-dimensional survey region 
  with cubic 4 $h^{-1}$ Mpc pixels ; we then count
  the number of galaxies within a radius of $r_{\rm ball}$
  centered on each pixel.
When we find $\leq 1$ galaxy 
  within $r_{\rm ball}$, the pixel is a void pixel.
We next connect the void pixels using a friends-of-friends algorithm
  to identify connected under-dense regions.
We expand the void only to a distance of $r_{\rm ball}-d_c$
  to account for a buffer region 
  neighboring the over-dense structures \citep{park12lss}.
The length $d_c$ is linking length 
  we use to identify over-dense structures. 

The right panels of Figure \ref{fig-num}
  show the number of under-dense structures
  we identify as a function of the ball radius, $r_{\rm ball}$.
The upper and lower panels show the numbers of structures
  with more than 30 and 20 connected void pixels, respectively
  (see filled circles for the HectoMAP data).
Similar to the over-dense case,
  the number of under-dense structures generally 
  first increases with ball radius, but
  decreases when the ball radius is large enough.
The change in the number of under-dense structures
  appears more sensitive to the change in effective linking length 
  than the number of over-dense structures (left panel).
We choose a ball radius of 
  $r_{\rm ball}=1.64d_{\rm mean}=$ 14.78 $h^{-1}$ Mpc 
  as the critical ball radius and a minimum of 20 pixels; 
  this radius yields the maximal number of under-dense regions.
We examine the impact of different ball radii
  in Section \ref{results}.

The bottom panel of Figure \ref{fig-hpid}
  highlights the under-dense structures with colored symbols.
Many of the voids are elongated with complex morphologies.
The largest under-dense region in both size and volume is 
  at R.A.$=$244.0 (deg), Decl.$=$43.3 (deg) and $z=0.37$;
  with a maximum extent of 300.8 $h^{-1}$ Mpc and
  a volume of $3.80\times10^5$ $h^{-3}$ Mpc$^3$.
The maximum extent of voids is the maximum separation of
 the void pixels in the structure. 
We compute this separation in real space.
The largest SDSS void complex in \citet{park12lss}
  has a volume of $1.44\times10^6$ $h^{-3}$ Mpc$^3$
  with a maximum extent of 334 $h^{-1}$ Mpc,
  apparently larger than the one in HectoMAP.
However, the ball radius used in Park et al.
  is 1.45 $d_{\rm mean}$ ($=$13.06 $h^{-1}$ Mpc)
  for the sample of SDSS galaxies 
  with $M_r\leq-21.6$, $z<0.17$, 
  and $d_{\rm mean}=9$ $h^{-1}$ Mpc.
If we use the same ball radius 
  as in Park et al. (i.e. $1.45 d_{\rm mean}$),
  the largest void in HectoMAP
  has a maximum extent of 883.6 $h^{-1}$ Mpc 
  with a volume of $2.01\times10^6$ $h^{-3}$ Mpc$^3$,
  larger than the largest void in the SDSS
  (\citealt{park12lss}, see also \citealt{pan12, sut12cat}
  for size distribution of SDSS voids).
As for the dense structures, 
  differences between HectoMAP and SDSS may result partly 
  from differences in survey geometry, but
  this exercise again emphasizes the necessity of
  using identical procedures when comparing physical properties of
  large-scale structure in different redshift surveys 
  or when comparing the data with a simulation.

\begin{figure}
\center
\includegraphics[width=0.48\textwidth]{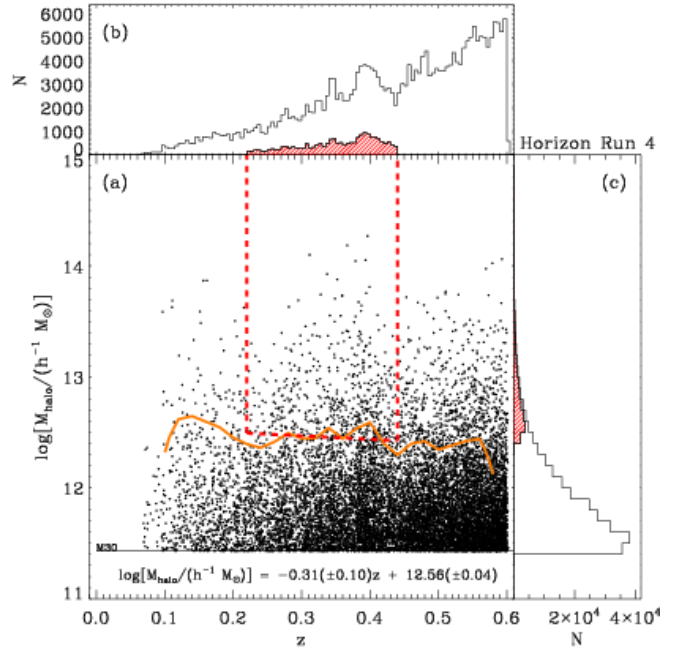}
\caption{Halo masses as a function of redshift in one mock sample
  from the Horizon Run 4 simulation.
The orange solid line is the lower limit for the halo sample that
  provides a constant halo number density over the redshift range 
  $0.1<z<0.57$ 
  ($n_{\rm halo}=1.36\times10^{-3}$ $h^{3}$ Mpc$^{-3}$ or 
   $d_{\rm mean}=9.01$ $h^{-1}$ Mpc).  
The slanted red dashed line is the best fit to the orange line, and 
  defines a volume-limited sample of halos delimited by the vertical dashed lines.
The horizontal solid line indicates 
  a minimum subhalo mass of 2.7$\times$10$^{11}$ $h^{-1}$ M$_\odot$
  with 30 dark matter particles.
We display only 5\% of the data for clarity.
The open and red hatched histograms in top and right panels
  show the distributions of halos 
  in the entire and volume-limited samples, respectively.
}\label{fig-mza}
\end{figure}

\section{ANALYSIS OF HORIZON RUN 4}\label{simiden}

Here we identify large-scale structures 
  in the Horizon Run 4 simulation data
  using the same procedure we use for the observations.
Sections \ref{ransamp} and \ref{densamp} describe two methods of
  sampling the halos from the mock surveys.
   
\subsection{Large-Scale Structures in Horizon Run 4: 
  Sampling Based on Halo Mass}\label{ransamp}

Figure \ref{fig-mza} shows the distribution of halo masses 
  as a function of redshift for one Horizon Run 4 mock survey.
As for the HectoMAP data (Figure \ref{fig-absz}),
  we first determine the lower mass limit that 
  gives a constant comoving number density of 
  $1.36\times10^{-3}$ $h^3$ Mpc$^{-3}$ at each redshift (orange solid line).
We then derive the best-fit 
  linear approximation to the contour (slanted red dashed line).
The red dashed lines define the volume-limited sample
  of halos we use for a statistical comparison
  with the HectoMAP data. 
The limiting redshifts are the same as for HectoMAP.

To take the effect of our small spectroscopic incompleteness
  into account in the simulated data,
  we compute the spectroscopic completeness
  at each redshift and absolute magnitude in Figure \ref{fig-absz}
  based on the completeness curve as a function of apparent magnitude
  (i.e. top panel of Figure \ref{fig-comp}).
Because the galaxy number density 
  at each redshift and absolute magnitude in Figure \ref{fig-absz}
  corresponds to the halo number density 
  at each redshift and halo mass in Figure \ref{fig-mza},
  we  remove halos in the appropriate bins of 
  the simulations to match the data.
By considering the spectroscopic completeness
  at each redshift and halo mass in Figure \ref{fig-mza},
  we ensure that the number and distribution of halos 
  we select from the simulations match
  number of observed galaxies 
  in the volume-limited sample (i.e. 9881 galaxies).

\begin{figure}
\center
\includegraphics[width=0.48\textwidth]{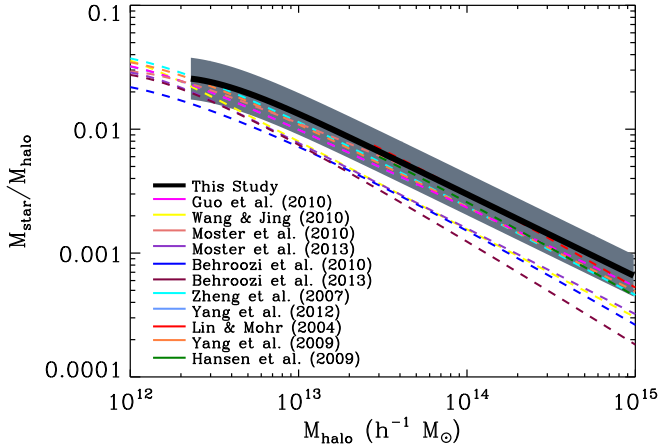}
\caption{Stellar-to-halo mass ratio as a function of halo mass
  for the volume-limited samples of 
  HectoMAP galaxies and Horizon Run 4 subhalos.
The black solid line is the best-fit relation  
  in the halo mass range 
  $2.3\times10^{12}-9.9\times10^{14}$ $h^{-1}$ $M_\odot$.
The gray shaded region indicates the 1$\sigma$ dispersion of the distribution 
  of galaxies and halos around the best-fit relation.  
For comparison, we show (colored dashed lines) relations 
  from other studies with different methods: abundance matching 
  \citep{guo10, wj10, mos10, mos13, beh10, beh13},
  the halo occupation distribution 
  \citep{zheng07},
  the  conditional luminosity function  
  \citep{yang12}, and
  cluster catalogs 
  \citep{lm04, yang09, han09}.
See \citet{beh13} for a more complete list of the relation 
  with a detailed discussion.
}\label{fig-mass}
\end{figure}
  
To show the relationship between the subhalos in Horizon Run 4 
  and the galaxies in HectoMAP,
  we plot the stellar-to-halo mass ratio
  as a function of halo mass in Figure \ref{fig-mass} 
  for the volume-limited samples.
We compute stellar masses 
  using the SDSS five-band photometric data with the 
  Le Phare\footnote{\url{http://www.cfht.hawaii.edu/~arnouts/lephare.html}} 
  code \citep{arn99,ilb06}.
Details of the stellar mass estimates are in \citet{zah14}. 
Because we use $r$-band absolute magnitudes
  to define the HectoMAP volume-limited sample (see Figure \ref{fig-absz}),
  we first sort the galaxies according to their $r$-band absolute magnitudes.
We then sort the subhalos by mass (see Figure \ref{fig-mza}),
  and associate them with the galaxies in HectoMAP sample
  according to their rank.
We combine all the subhalo-galaxy association from the 300 mock surveys, and
  fit to the data with the functional form in \citet{mos10,mos13},
\begin{equation}
\frac{M_{\rm star}}{M_{\rm halo}} = 
  2 N {\left[ {\left( M_{\rm halo}/M_1  \right)}^{-\beta} + 
  {\left( M_{\rm halo}/M_1  \right)}^{\gamma} \right]}^{-1}.
\end{equation}
Because we cannot provide a good constraint on the slope at a small mass range 
  (i.e. $M_{\rm halo}<10^{12}$ M$_\odot$),
  we fix $\beta=1$ \citep{mos10,mos13}.
The best-fit parameters for the halo mass range 
  $2.3\times10^{12}-9.9\times10^{14}$ $h^{-1}$ M$_\odot$ are 
  $N=0.026$, $M_1=1.38\times10^{12}$ $h^{-1}$ M$_\odot$, and $\gamma=0.664$.
We plot the best-fit relation in Figure \ref{fig-mass} as a black solid line;
  the stellar-to-halo mass ratio decreases with increasing subhalo mass.
The gray shaded region indicates the 1$\sigma$ dispersion of the distribution 
  of the data around the best-fit relation. 
For comparison, we show the relations (colored dashed lines) 
  from other studies with different methods: 
  abundance matching (e.g., \citealt{guo10}), 
  the halo occupation distribution (e.g., \citealt{zheng07}), 
  the conditional luminosity function (e.g., \citealt{yang12}), 
  and  cluster catalogs (e.g., \citealt{lm04}).
This comparison substantiates our approach; 
  our relation is consistent with a range of previous studies.
There are several effects that 
  may contribute to the large scatter among the relations
 (e.g., different simulation resolution, 
  large uncertainty in stellar mass estimates,
  different galaxy samples, or different halo identification methods), but 
  a detailed discussion is beyond the scope of this paper
 (see \citealt{beh13} for a more complete list of the relation 
  with a detailed discussion).

\begin{figure*}[ht]
\center
\begin{tabular}{c}
\includegraphics[width=0.99\textwidth]{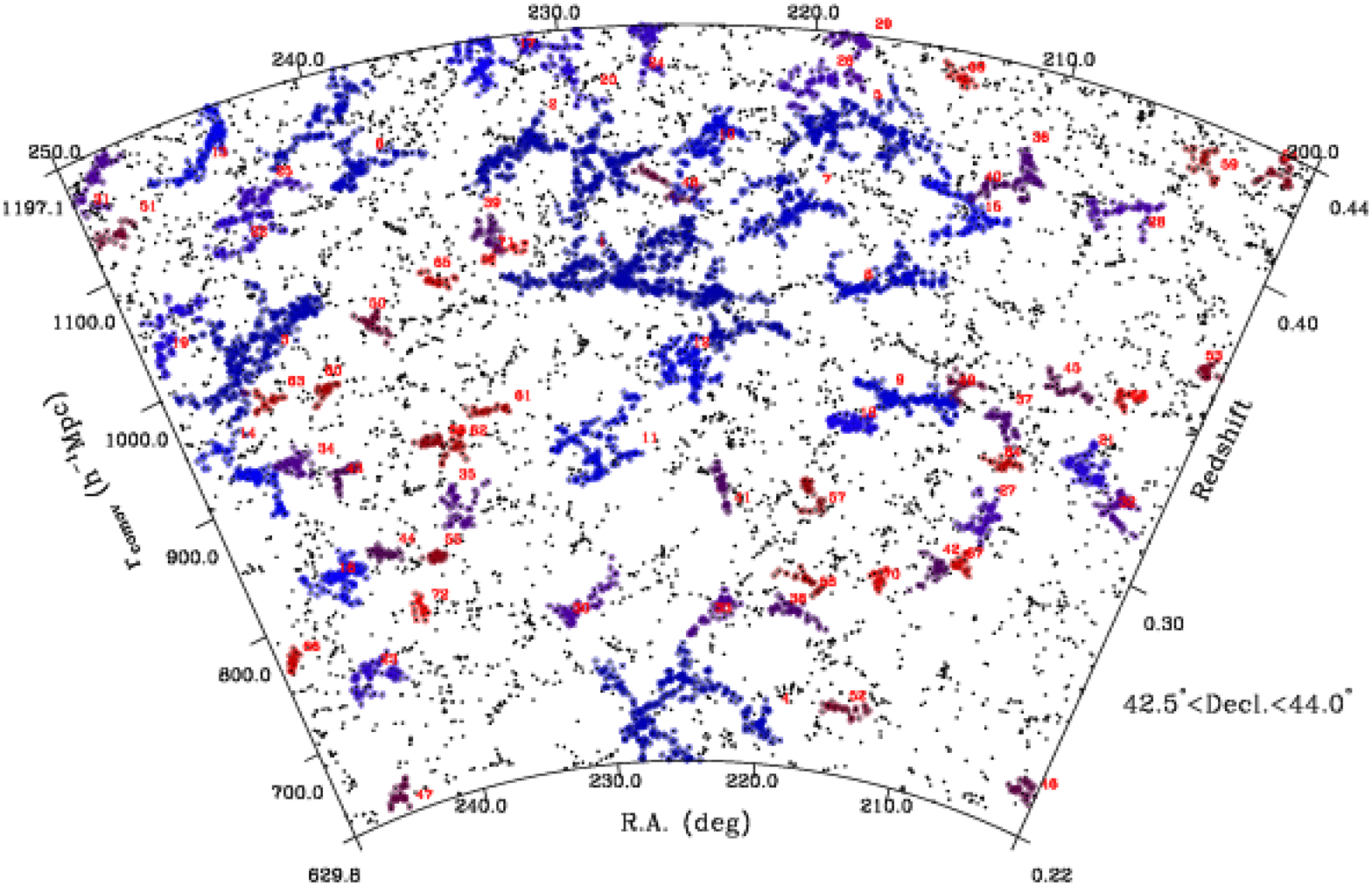} \\
\includegraphics[width=0.99\textwidth]{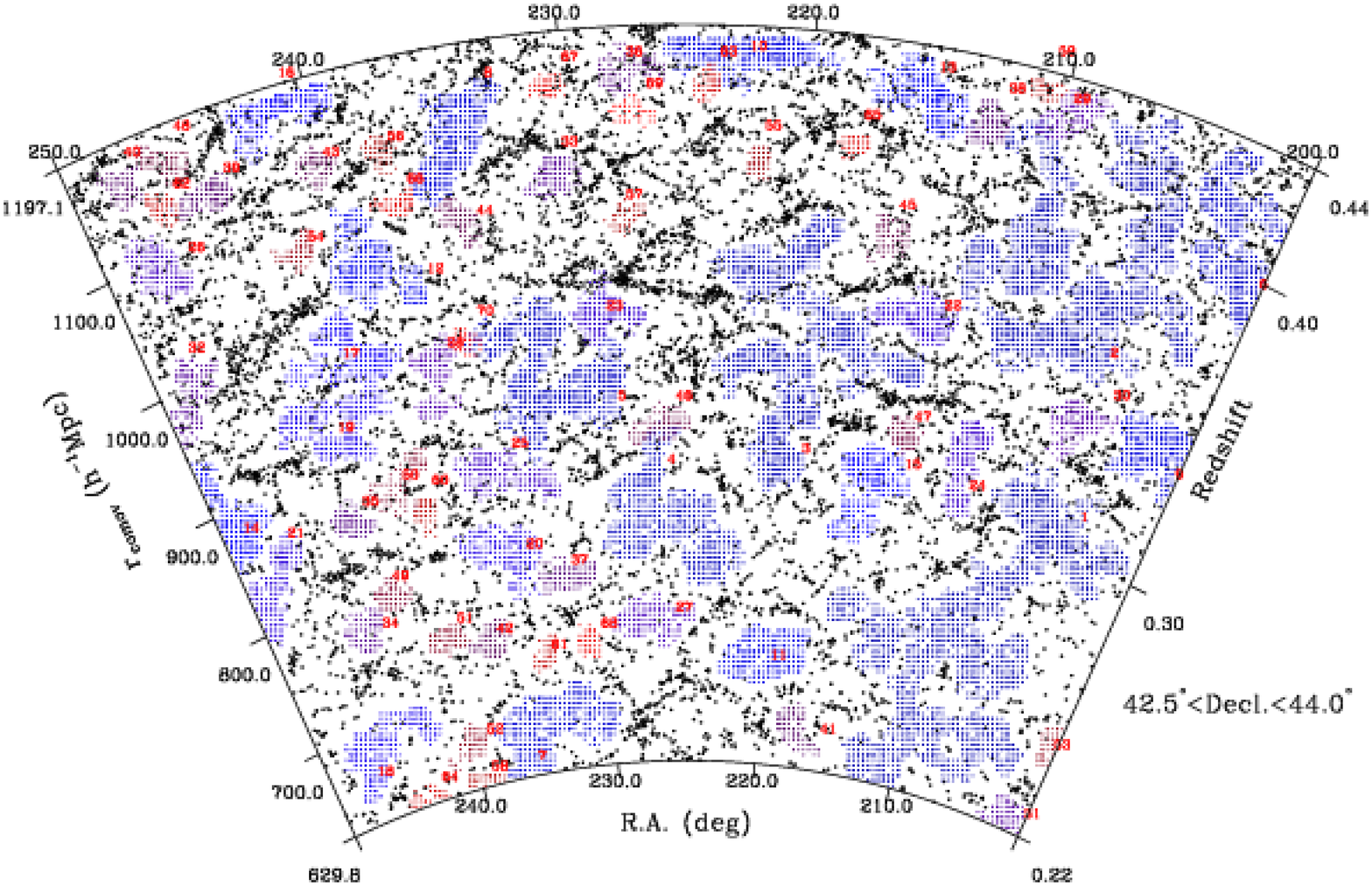}
\end{tabular}
\caption{Same as Figure \ref{fig-hpid},
  but for one of the mock surveys from Horizon Run 4.
}\label{fig-hr4a}
\end{figure*}

We now apply the friends-of-friends algorithm 
  to the volume-limited samples
  of halos from the 300 mock surveys to identify 
  over-dense large-scale structure.
We also reduce fingers as in the HectoMAP data.
The left panels of Figure \ref{fig-num} (open circles) show 
  the number of over-dense structures from 300 mock surveys
  as a function of linking length (mean at each linking length).
The overall behavior is similar to the one 
  for the HectoMAP data.
Using the same critical linking length as for the HectoMAP data,
  $d_c=0.9d_{\rm mean}$,
  we show the spatial distribution of structures
  marked by member halos in the top panel of Figure \ref{fig-hr4a}.  
The structures are similar to the over-dense structures in HectoMAP
  (Figure \ref{fig-hpid}).
 
To identify under-dense features in the simulation,
  we again tessellate the three-dimensional survey region 
  with cubic pixels,
  and count the number of halos within a radius of $r_{\rm ball}$
  centered on each pixel.
By connecting the void pixels with $\leq 1$ halo 
  inside  $r_{\rm ball}$,
  we obtain a list of under-dense regions.

The right panels of Figure \ref{fig-num}
  show the number of simulated under-dense structures 
  (open circles)
  as a function of ball radius, $r_{\rm ball}$.
Again, the numbers of under-dense regions
  in the simulation and observations behave similarly.
We adopt the same ball radius of $r_{\rm ball}=1.64d_{\rm mean}$
  as for HectoMAP because of the known sensitivity 
  of the structures 
  we identify to this scale (see the discussion 
  of the comparison of
  the largest HectoMAP structures with the SDSS results
  at the end of Sections \ref{overden} and \ref{underden}).
The bottom panel of Figure \ref{fig-hr4a} shows
  the Horizon Run 4 under-dense regions
  from one mock survey.
Again, the morphologies are diverse and often complex.
Section \ref{results} contains a further discussion 
  of the physical properties of these voids.

\begin{figure*}
\center
\begin{tabular}{cc}
\includegraphics[width=0.47\textwidth]{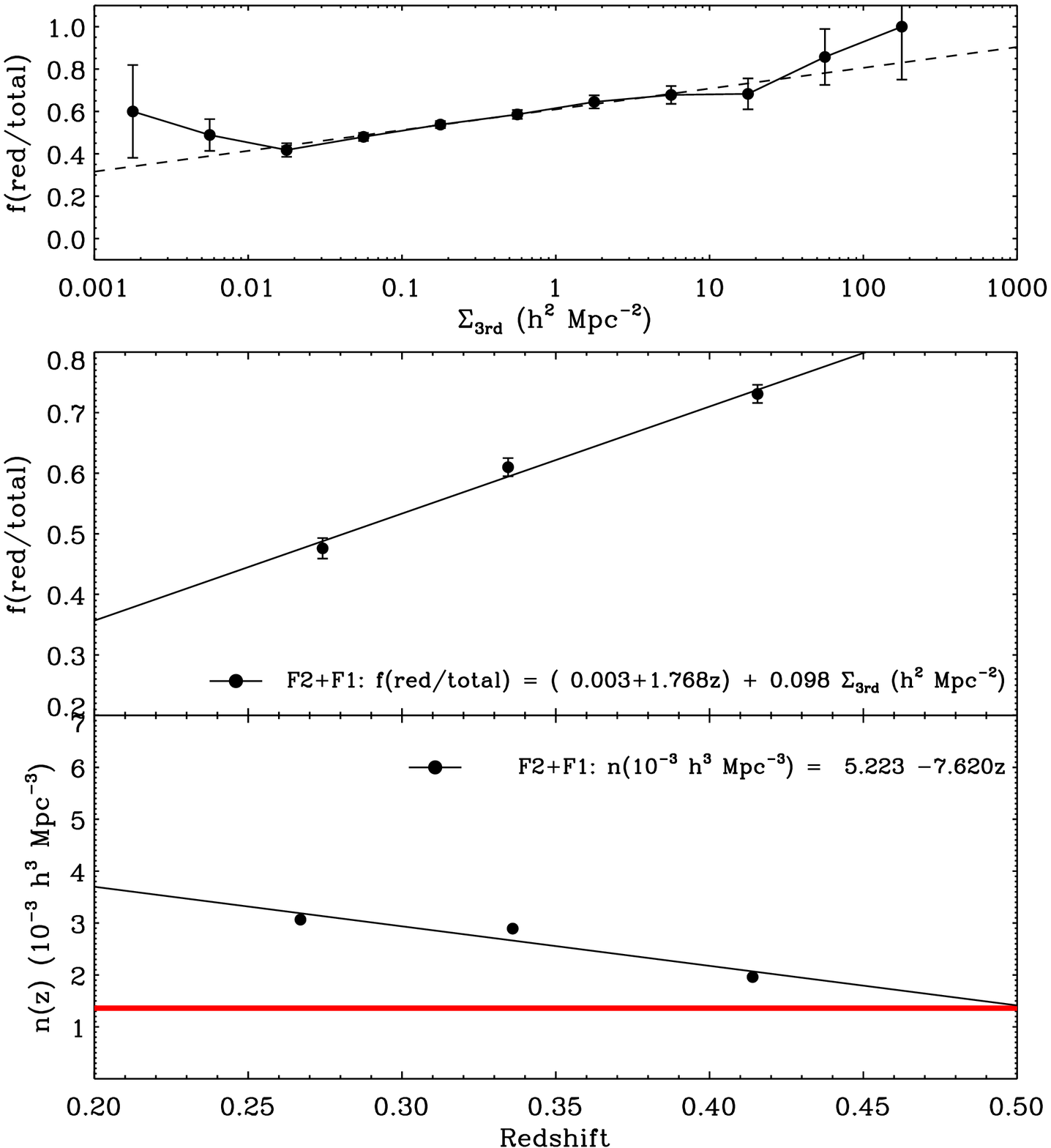}
\includegraphics[width=0.51\textwidth]{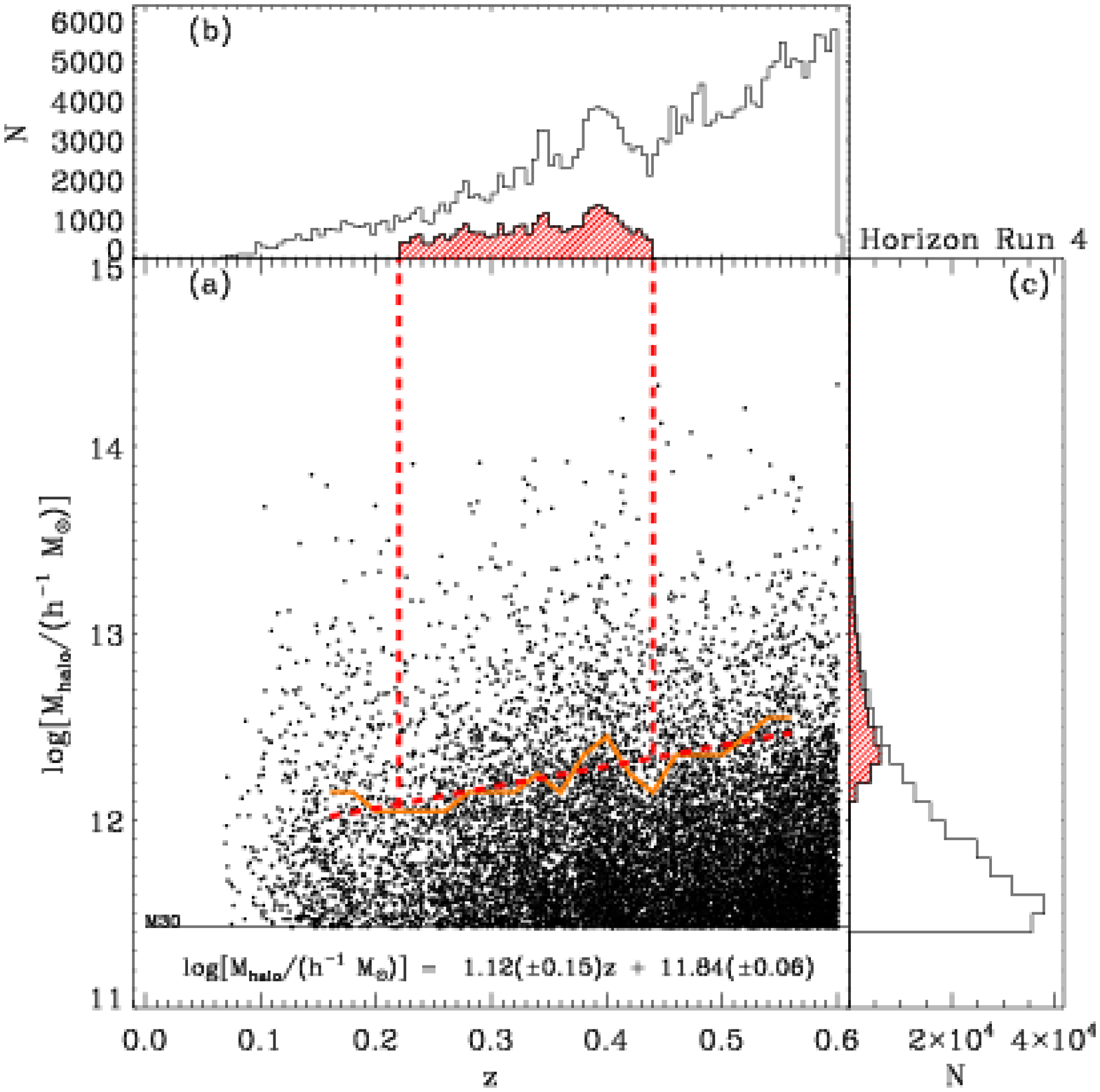}
\end{tabular}
\caption{({\it Bottom left}) Galaxy number density 
  in the combined data of F1 and F2
  as a function of redshift
  using all the galaxies regardless of color (filled circles).
The black solid line is the best-fit to the data.
The red solid line indicates a constant number density
  of galaxies satisfying the HectoMAP color selection.
({\it Top left}) Fraction of HectoMAP selected galaxies as a function of 
  surface galaxy number density
  in the combined F1 and F2 data at $0.22<z<0.44$.
The dashed line is the best fit to the data.
({\it Middle left}) Change in HectoMAP selected galaxy fraction 
  as a function of redshift
  for the combined F1 and F2 data (filled circles).
The black solid line is the best-fit to the data.
({\it Right}) Same as Figure \ref{fig-mza},
  but for one mock survey from  Horizon Run 4.
Note that the sampling of halos here is based on
  local density; it differs from the halo sample 
  in Figure \ref{fig-mza}.
We display only 5\% of the data for clarity.
}\label{fig-mzb}
\end{figure*}

\subsection{Horizon Run 4 Large-Scale Structures: 
Sampling Halos Based on Local Density}\label{densamp}

\subsubsection{Sampling Halos}

When we construct a volume-limited sample of halos 
  in Section \ref{ransamp},
  we match the number density of observed galaxies 
  with that of halos by assuming that 
  more luminous massive halos correspond to more luminous galaxies
  without considering the HectoMAP red selection in detail. 
Although this approach is acceptable 
  because the red galaxies are generally good tracers 
  of matter distribution (e.g., \citealt{mad03,park07,zeh11,kurtz12}),
  here we use local densities around halos
  to mimic the HectoMAP red selection.
To evaluate the density distribution 
  around the observed galaxies in HectoMAP, 
  we use the SHELS (Smithsonian Hectospec Lensing Survey) survey. 
SHELS covers two  
  separate 4 square degree fields of the Deep Lens Survey (F1 and F2)
  \citep{wit02,wit06}.
Currently SHELS is the densest redshift survey to its limiting
  apparent magnitude, $R=20.6$ (or $r\sim20.9$).
Unlike HectoMAP, SHELS is
  a complete magnitude-limited survey 
  \citep{gel05,gel10,gel14f2,gel15f1}. 
In other words, there is no color selection. 
Thus it can be used as a base 
  for examining the HectoMAP color selection \citep{gh15}.

We use the complete magnitude-limited survey, SHELS, 
  to compute the fraction of red galaxies 
  satisfying the HectoMAP selection
  as a function of surrounding local density and redshift.
We then apply the relation we determine from SHELS
  to the simulation to select the fraction of halos
  that mimics the fraction of red galaxies 
  with a given surrounding local density and redshift.

The steps in our construction of a sample of halos 
  matching the red selection of HectoMAP are:

\begin{enumerate}

\item[(1.1)] Using SHELS,
  we first construct a volume-limited sample of galaxies 
  with $0.22<z<0.44$, 
  similar to the sample in HectoMAP.
We define the absolute magnitude limit as a function of redshift
  so that the number density of galaxies satisfying 
  the HectoMAP color selection  is constant with redshift as
   in  HectoMAP  (i.e. $d_{\rm mean}=9.01$ $h^{-1}$ Mpc).
Because the fraction of HectoMAP selected galaxies changes with redshift,
  the number density of all the galaxies regardless of color
  in this volume-limited sample changes with redshift;
  the bottom left panel in Figure \ref{fig-mzb} shows this change.
  
\item[(1.2)] Next we compute the surface galaxy number density 
  around each galaxy using all the galaxies regardless of color  
  with relative velocities $ \Delta v \leq 1000$ km s$^{-1}$. 
We compute $\Sigma_3=3(\pi D^2_{p,3})^{-1}$ 
  where $D_{p,3}$ is the projected distance 
  to the 3rd-nearest neighbor galaxy.
The typical physical scale of this $\Sigma_3$ probes is 
  $2-3$ $h^{-1}$ Mpc, and our results do not change 
  even if we use 5th-nearest neighbor galaxy.

\item[(1.3)] We compute the fraction of HectoMAP selected galaxies 
  as a function of $\Sigma_3$ in three redshift bins 
  ($0.22<z<0.3$, $0.3<z<0.37$, and $0.37<z<0.44$),
  and determine the best-fit relation between this
  fraction and $\Sigma_3$ at each redshift:
  f(red/total)$= a + b \Sigma_3$.
Figure \ref{fig-mzb} (upper left panel) shows an example
  of this fit at $0.22<z<0.44$.

\item[(1.4)] We combine the best-fit relations 
  in the three redshift bins to determine a global relation 
  among the red galaxy fraction, redshift and local density:
  f(red/total)$= a_0 + a_1 z + b \Sigma_3$ where
  $a_0=0.003\pm0.058$, $a_1=1.768\pm0.158$ and 
  $b=0.098\pm0.013$
  are the coefficients derived for the best-fit relation.
Figure \ref{fig-mzb} (middle left panel) shows 
  the redshift dependence of this relation 
  when $b=0$ as an example. 

\end{enumerate}

\begin{figure*}[ht]
\center
\begin{tabular}{c}
\includegraphics[width=0.99\textwidth]{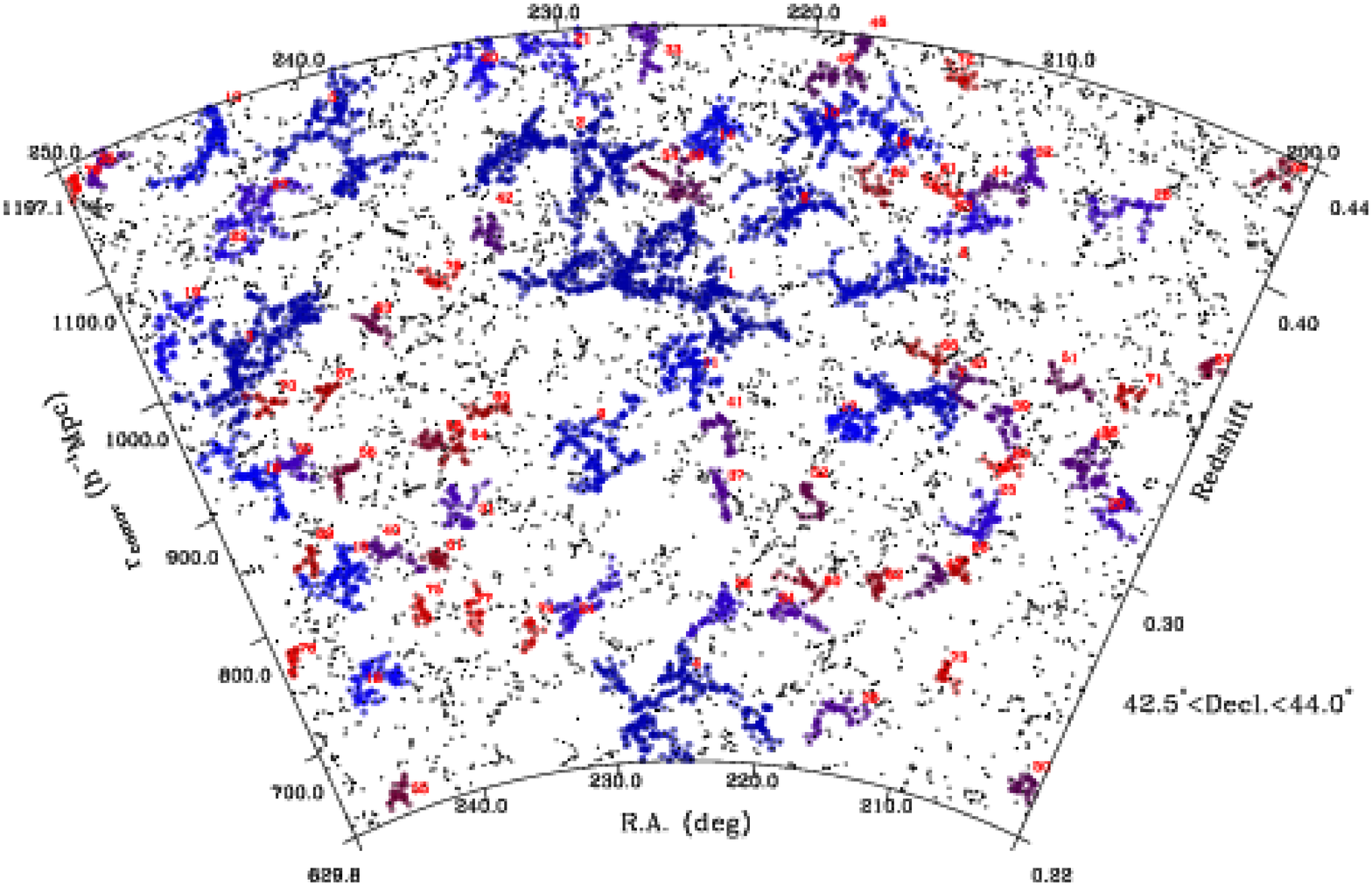} \\
\includegraphics[width=0.99\textwidth]{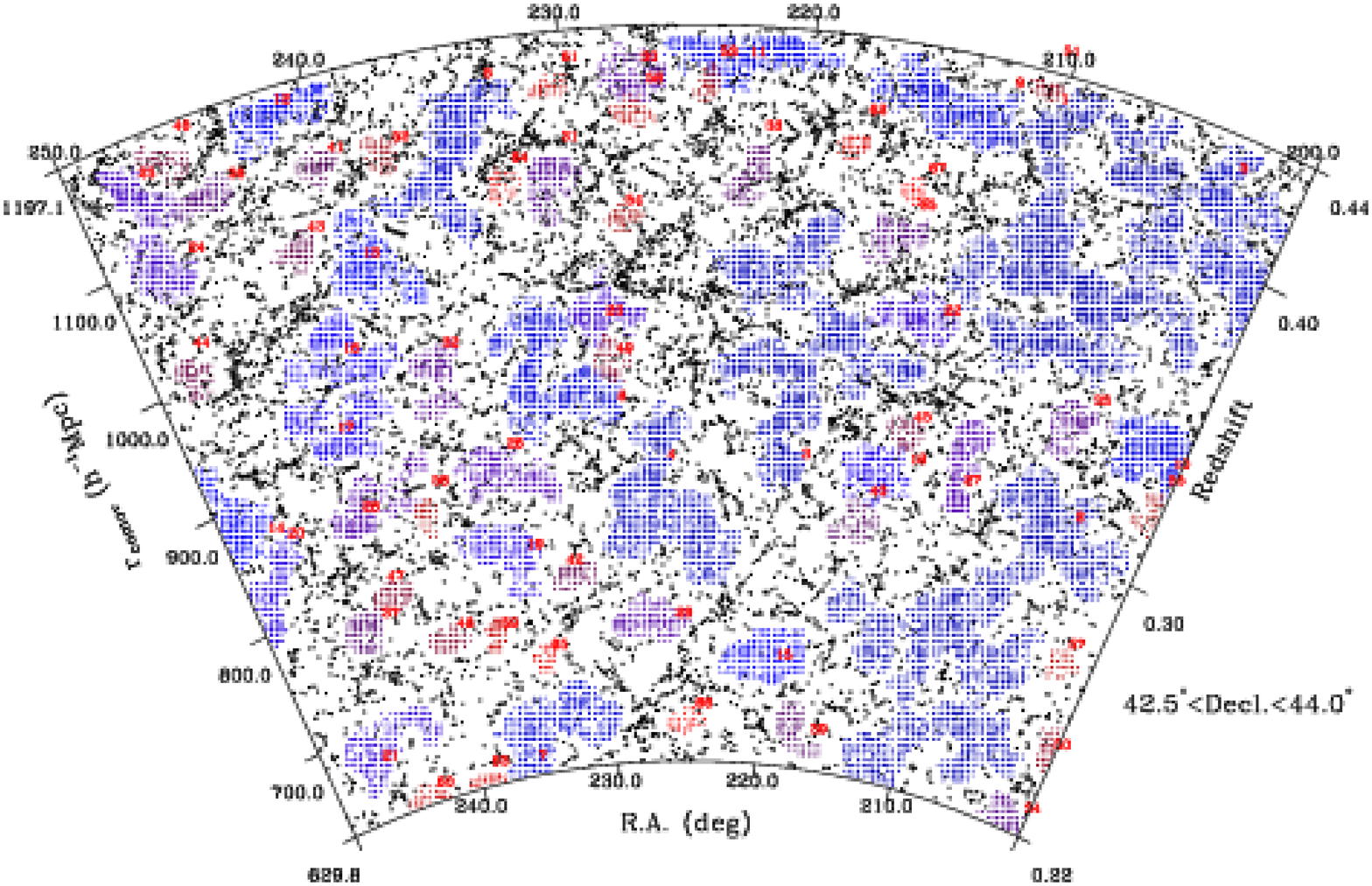}
\end{tabular}
\caption{Same as Figure \ref{fig-hpid},
  but for one of the mock surveys from Horizon Run 4.
The sampling of halos here is based on
  local density; it differs from the halo sample 
  in Figure \ref{fig-hr4a}.
}\label{fig-hr4b}
\end{figure*}

For a  galaxy sample following the relation,
  f(red/total)$= 0.003 + 1.768 z + 0.098 \Sigma_3$, the
  number density of red galaxies is independent of redshift
  as in the volume-limited sample of HectoMAP galaxies.
We thus apply this relation
  to the Horizon Run 4 mock surveys next.

\begin{figure*}
\center
\includegraphics[width=0.99\textwidth]{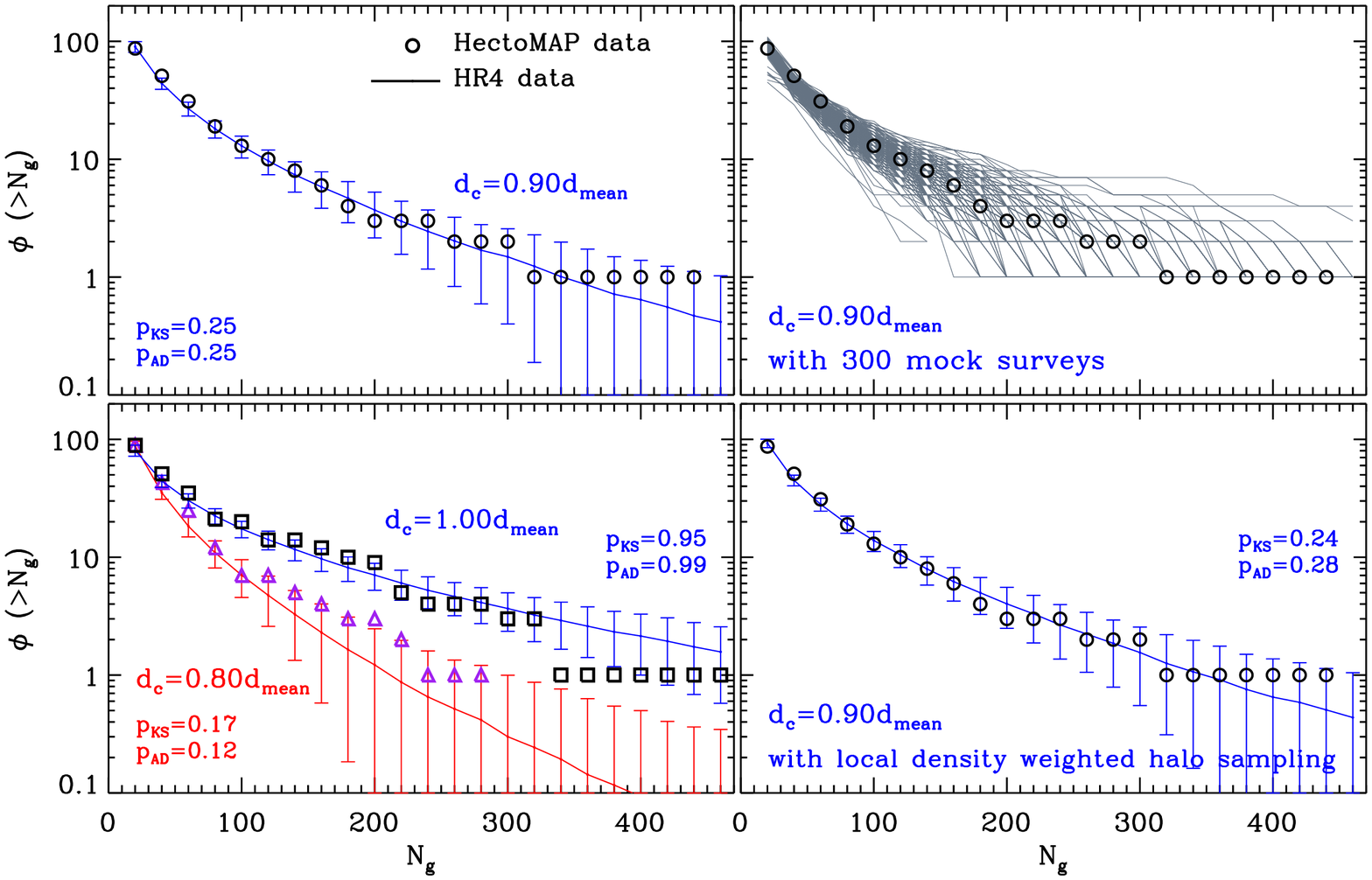}
\caption{({\it Top left}) Cumulative richness distribution
  of over-dense large-scale structures identified with the critical linking length
  of  $d_c=0.9d_{\rm mean}$ 
  (HectoMAP: open circles, Horizon Run 4: blue solid lines).
The error bars indicate the standard deviation
  in the richness distributions from 
  the 300 mock surveys of Horizon Run 4.
Two numbers in the left corner indicate $p$-values
  from the K-S and A-D k-sample tests on the distributions
  of the HectoMAP and Horizon Run 4 data.
The sampling of halos is based on halo mass (see Section \ref{ransamp}).
The gray curve in the top right panel indicates the distribution 
  for each mock survey,
  and the open circles are the same as in the left panel.
({\it Bottom left}) Same as top left panel,
  but for the distributions with two different linking lengths
  ($d_c=1.0d_{\rm mean}$: squares and blue solid line,
   $d_c=0.8d_{\rm mean}$: triangles and red solid line).
({\it Bottom right}) Same as top left panel,
  but for the distribution from different mock surveys of the Horizon Run 4,
  based on local density weighted halo sampling (see Section \ref{densamp}).
}\label{fig-den}
\end{figure*}

\begin{figure*}
\center
\includegraphics[width=0.99\textwidth]{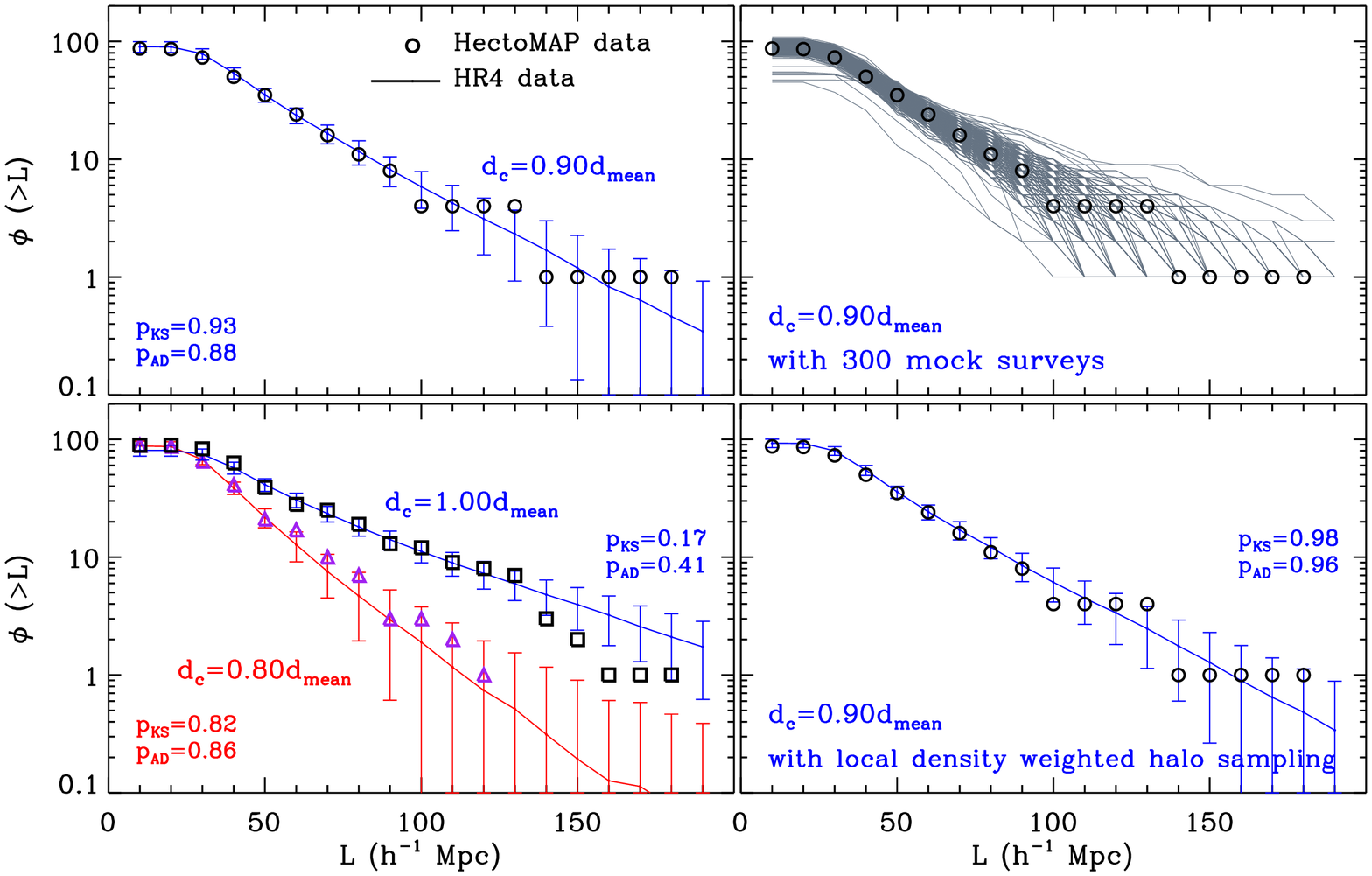}
\caption{Same as Figure \ref{fig-den}, but for
  the cumulative size distribution
  of over-dense large-scale structures.
}\label{fig-densize}
\end{figure*}

\begin{enumerate}
    
\item[(2.1)] In the plot of halo mass and redshift for  a mock survey,
  we determine the lower mass limit 
  that follows the change of galaxy number density with redshift 
  determined at (1.1); 
  the orange line in the right panel of Figure \ref{fig-mzb} 
  shows this limit.
The best-fit relation to the orange line (slant red dashed line) 
  defines the volume-limited sample of halos at $0.22<z<0.44$.

\item[(2.2)] With the volume-limited sample of halos defined at (2.1), 
  we compute the surface number density  around each halo
  using the 3rd-nearest neighbor halo:
  $\Sigma_3=3(\pi D^2_{p,3})^{-1}$.

\item[(2.3)] We select a fraction of halos
  at each redshift and local density from the mock survey
  that matches the fraction of HectoMAP selected galaxies 
  determined from the relation 
  among the HectoMAP selected galaxy fraction, redshift and local density at (1.4).
We keep the total number of selected halos in a mock survey
  the same as the total number 
  of HectoMAP galaxies in the volume-limited sample
  at $0.22<z<0.44$ (i.e. $N=9881$).
\end{enumerate}

\subsubsection{Horizon Run 4 Large-Scale Structure with Local Density Selection}

We again apply the friends-of-friends algorithm 
  to the 300 volume-limited samples of halos 
  selected to mimic the observational selection. 
We identify over-dense large-scale structures.
Figure \ref{fig-hr4b} (upper panel) shows 
  the spatial distribution of structures marked by these halos
  in one mock sample based with
  a linking length of $d_c=0.9d_{\rm mean}$.
The bottom panel shows the distribution 
  of under-dense regions using the void-finding algorithm 
  of Section \ref{underden}.
Both the over-dense and under-dense structures in the
  simulations and observations are indistinguishable by eye. 
We compare the simulations and the observations 
  quantitatively in the next section.

\section{COMPARISON OF OBSERVED AND SIMULATED LARGE-SCALE STRUCTURES}\label{results}

Here we make a quantitative comparison between the
  over-dense and under-dense structures
  in  HectoMAP and the Horizon Run 4 simulations.
We first compare over-dense structures, and then under-dense structures. 
We use several measures to compare distributions of the properties of these structures. 
For the non-parametric measures, 
  the Kolmogorov-Smirnov (K-S) test and the Anderson-Darling (A-D) k-sample test, 
  we list the relevant $p$-value in each figure. 
The $p$-values indicate the probability that the data and the simulations 
  are drawn from the same parent distribution.

The physical properties of the observed large-scale structures
   depend strongly on (or are restricted by) the survey geometry.
The HectoMAP is a thin slice survey, 
  covering a 53 deg$^2$ strip that is only 1.5 deg wide.
Although the extent of a structure along the declination direction 
  is essentially unconstrained by the data,
  there is no bias in our comparison 
  between observations and simulations
  because the Horizon Run 4 mock surveys have the same
  geometry as HectoMAP.

\subsection{Over-dense Large-Scale Structures}\label{overlss}

Figure \ref{fig-den} (upper left) shows 
  the cumulative richness distribution of over-dense structures 
  in  HectoMAP (open circles) and in the Horizon Run 4 simulation
  (blue solid curve). More precisely, the cumulative distribution is the
  number of structures per survey volume 
  ($7.89\times10^6$ $h^{-3}$ Mpc$^3$) 
  that include $\geq n_g$ member galaxies.
The critical linking length is $d_c=0.9d_{\rm mean}$.
The error bar indicates the dispersion
  in the richness distribution based on the 300 mock surveys.
The plot shows that the richness distributions
  of over-dense structures in both the observations
  and simulations data agree within the 1$\sigma$ error bars.

To highlight the richness distribution 
  from the 300 mock surveys,
  we show the distribution for each mock survey
  with a gray curve in the top right panel.
The distribution of the observations (open circles)
  lies well within the range of the distributions
  for the 300 mock surveys.

To examine the sensitivity of our results to the linking length,
  we show the richness distributions
  for the observations and simulations
  based on two different linking lengths (bottom left panel).
The differences between the observations
  and simulations, particularly for the $d_c=0.8d_{\rm mean}$, 
  appear to be slightly larger than for $d_c=0.9d_{\rm mean}$
  but the observations still agree with the simulations 
  within the error bars.

The bottom right panel shows the analogous richness distribution for
  the simulation data  based on local density sampling and 
  $d_c=0.9d_{\rm mean}$ (see Section \ref{densamp}).
The correspondence between the observations and the simulations
  is excellent. 

In all panels of Figure \ref{fig-den}, 
  the K-S and A-D $p$-values reject the null hypothesis
  at the $\lesssim 1.5\sigma$ level.
This weak rejection is consistent 
  with the comparison of the data and simulations based on the error bars.
  
Figure \ref{fig-densize} shows the cumulative size distribution
  of over-dense structures 
  (i.e. number of structures with maximum extent larger than $L$).
As in the previous plots,
  the top left panel shows the distributions
  of structures in the HectoMAP (open circles) and Horizon Run 4 data
  (blue solid curve) based on the critical linking
  length of $d_c=0.9d_{\rm mean}$.
The two distributions agree well.

\begin{figure*}
\center
\includegraphics[width=0.99\textwidth]{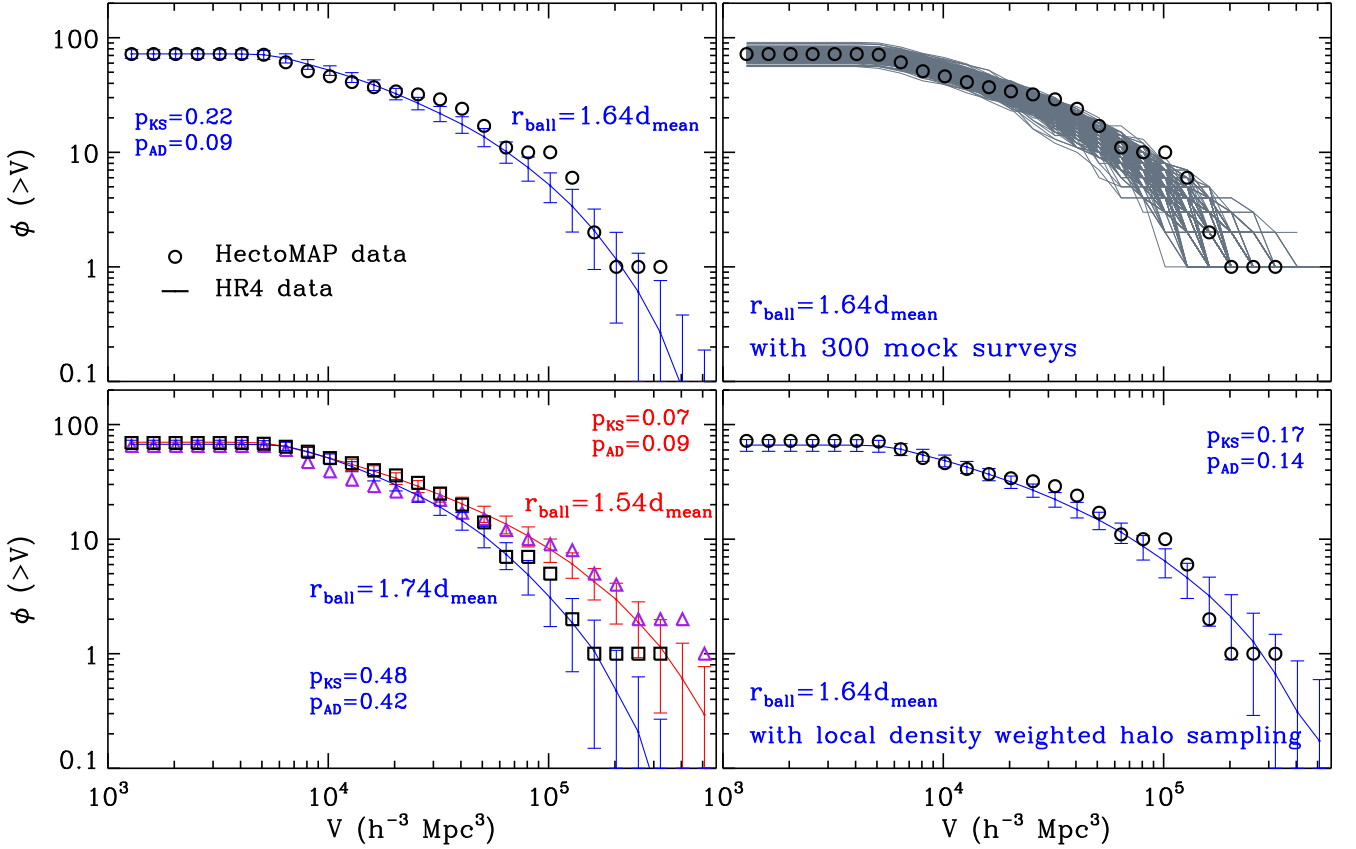}
\caption{({\it Top left}) Cumulative volume distribution
  of under-dense large-scale structures identified with the critical ball radius
  of $r_{\rm ball}=1.64d_{\rm mean}$ 
  (HectoMAP: open circles, Horizon Run 4: blue solid lines).
The error bars indicate the standard deviation
  in the size distributions from 
  the 300 mock surveys of  Horizon Run 4.
Two numbers in the left corner indicate $p$-values
  from the K-S and A-D k-sample tests on the distributions
  of the HectoMAP and Horizon Run 4 data.
The sampling of halos is based on halo mass (see Section \ref{ransamp}).
The gray curve in the top right panel indicates the distribution 
  from each mock survey,
  and the open circles are the same as in the top left panel.
({\it Bottom left}) Same as top left panel,
  but for  distributions with two different ball radii
  ($r_{\rm ball}=1.74d_{\rm mean}$: squares and blue solid line,
   $r_{\rm ball}=1.54d_{\rm mean}$: triangles and red solid line).
({\it Bottom right}) Same as top left panel,
  but for the distribution from different mock surveys of the Horizon Run 4,
  based on local density weighted halo sampling (see Section \ref{densamp}).
}\label{fig-void}
\end{figure*}

The top right panel compares the distribution
  from the observations (open circles) with 
  the 300 mock surveys  (gray curves).
The bottom left panel shows the result of using different
  linking lengths
  (i.e. $d_c=0.8d_{\rm mean}$ and $1.0d_{\rm mean}$).
The bottom right panel shows the impact of sampling the
  simulation based on local density sampling of halos
  (see Section \ref{densamp}).
All panels show that 
  the observed and simulated universes are remarkably similar. 
The K-S and A-D $p$-values reject the null hypothesis 
  for these distributions at an even weaker level 
  than for the distributions in Figure \ref{fig-den}.
Because the match between the two is excellent, 
  the dispersion in the 300 mock surveys can give
  us a robust measure of the probable error
  in the physical properties of observed large-scale structures.

\subsection{Voids: Under-dense Structure}\label{underlss}

Here we compare the volume and size distributions
  of under-dense large-scale structures (voids)
  in HectoMAP and the Horizon Run 4 simulation.
Figure \ref{fig-void} (top left panel)
  shows the cumulative volume distribution
  of under-dense structures in HectoMAP (open circles) and 
  in the Horizon Run 4 simulation (blue solid curve). 
The cumulative distribution tracks the number of structures with volume $> V$.
We use a critical ball radius, $r_{\rm ball}=1.64d_{\rm mean}$.
The error bar indicates the dispersion
  among the 300 mock surveys. 
The two distributions agree well even though
  there is a small offset between the two around $V\sim10^5$ ($h^{-3}$ Mpc$^3$).
The distributions from the individual 300 mock surveys 
  (gray curves in the top right panel)
  suggest that the small offset is statistically insignificant;
The planned deeper HectoMAP survey with $r_{\rm Petro,0}<21.3$ will 
  roughly double the galaxy number density.
With this denser sample, we will be able to study any systematic issue
  that might be responsible for this small difference.

The bottom left panel shows the volume distribution
  of under-dense structures 
  based on two different ball radii
 ($r_{\rm ball}=1.74d_{\rm mean}$: squares and blue solid line,
  $r_{\rm ball}=1.54d_{\rm mean}$: triangles and red solid line).
The distributions of observations and simulations 
  agree well within the error bars.
The bottom right panel shows a similar volume distribution
  with  $r_{\rm ball}=1.64d_{\rm mean}$, but for
  simulated halos based on local density sampling
  (see Section \ref{densamp}).
The correspondence between the observations and simulations
  appears better than the case
  based on halo mass sampling (top left panel).

In the panels of Figure \ref{fig-void}, 
  the K-S and A-D $p$-values reject the null hypothesis 
  at a significance $\lesssim 1.8 \sigma$. 
This weak rejection is again consistent with the correspondence
  between observations and simulations based on the error bars.
  
\begin{figure*}
\center
\includegraphics[width=0.99\textwidth]{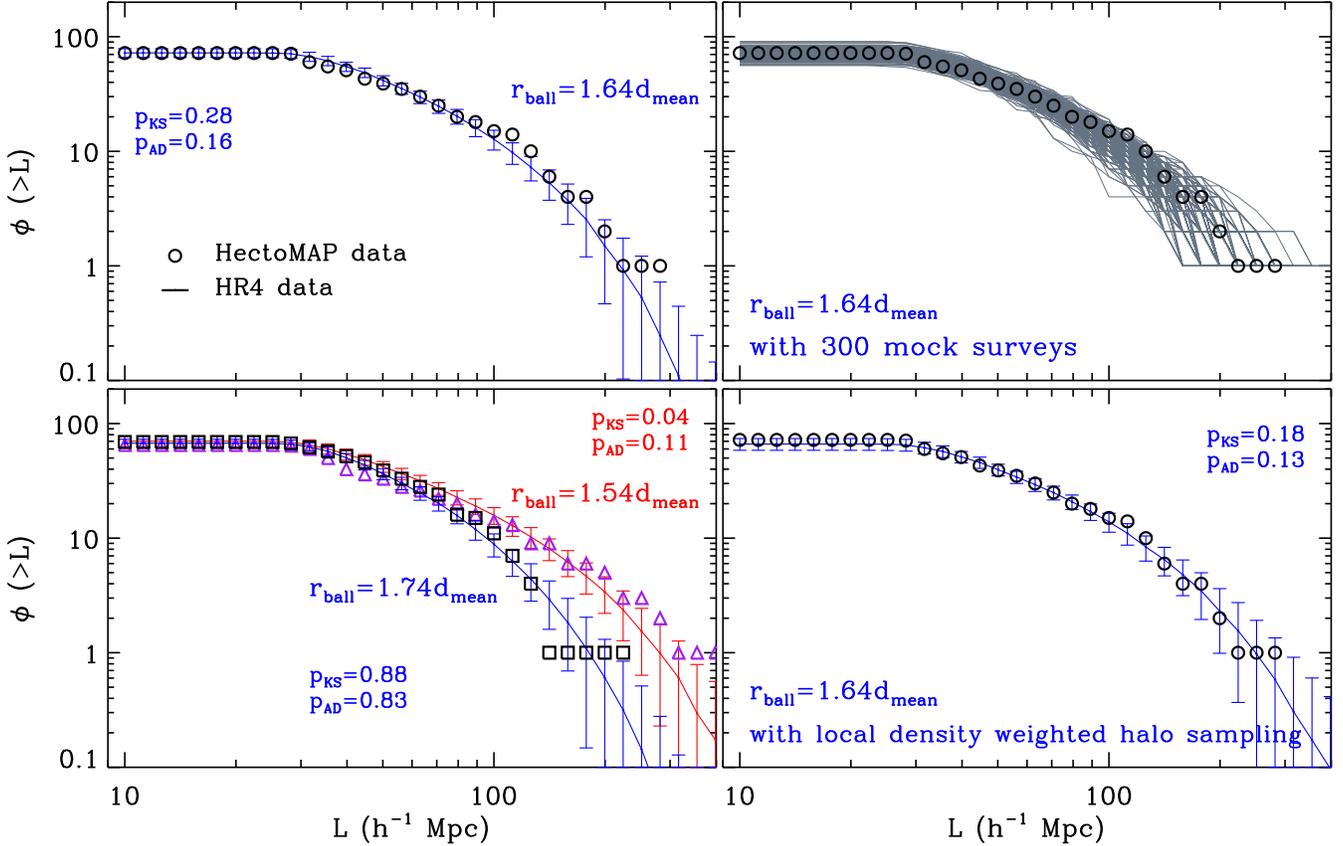}
\caption{Same as Figure \ref{fig-void}, but for
  the cumulative size distribution
  of under-dense large-scale structures.
}\label{fig-voidsize}
\end{figure*}

Figure \ref{fig-voidsize}
   shows the cumulative size distribution
  of under-dense structures 
  (i.e. number of structures with maximum extent  $ > L$).
  The top left panel shows the agreement between the distribution
  of HectoMAP (open circles) and Horizon Run 4 data
  (blue solid curve) based on 
  a critical ball radius of $r_{\rm ball}=1.64d_{\rm mean}$.  

The top right panel shows the distribution
  from the observations compared with the 300 mock surveys. 
The bottom left panel shows results for 
  two different ball radii 
  (i.e. $r_{\rm ball}=1.54d_{\rm mean}$ and $r_{\rm ball}=1.74d_{\rm mean}$). 
Here we see the only case where one of the statistical tests (the K-S test) 
  rejects the null hypothesis at the $2\sigma$ level. 
Even this rejection, limited to only one of the two tests, is weak.

The bottom right panel shows the result
  for the simulations with  halos selected 
  on the basis of local density sampling
  (see Section \ref{densamp}). 
Here the K-S and A-D statistics reject the null hypothesis 
  at the $\lesssim 1.5\sigma$ level.
The correspondence between the observations and simulations 
  is impressive in all of the comparisons we make.

\section {THE LARGEST STRUCTURES IN HectoMAP and HORIZON RUN 4} \label{big}

The largest structures known in the universe,
  the CfA Great Wall \citep{gh89} and the Sloan Great Wall \citep{gott05},
  led to interesting tests
  of models of structure formation.
Several studies asked
  whether the existence of such structures is compatible with
  a universe that is homogeneous and isotropic 
  on large scales (e.g., \citealt{clo12,hor13}).
This issue has been tested and resolved several times
  through careful statistical comparison of
  the physical properties of large-scale structure 
  in the observations and simulations, 
  particularly in the nearby universe
 \citep{spr06,sd11,park12lss,park15lss,alp14}. 
Here we revisit the issue by comparing the properties 
  of the largest structures in HectoMAP 
  with results from the 300 Horizon Run 4 mock surveys.

\begin{figure*}
\center
\includegraphics[width=0.97\textwidth]{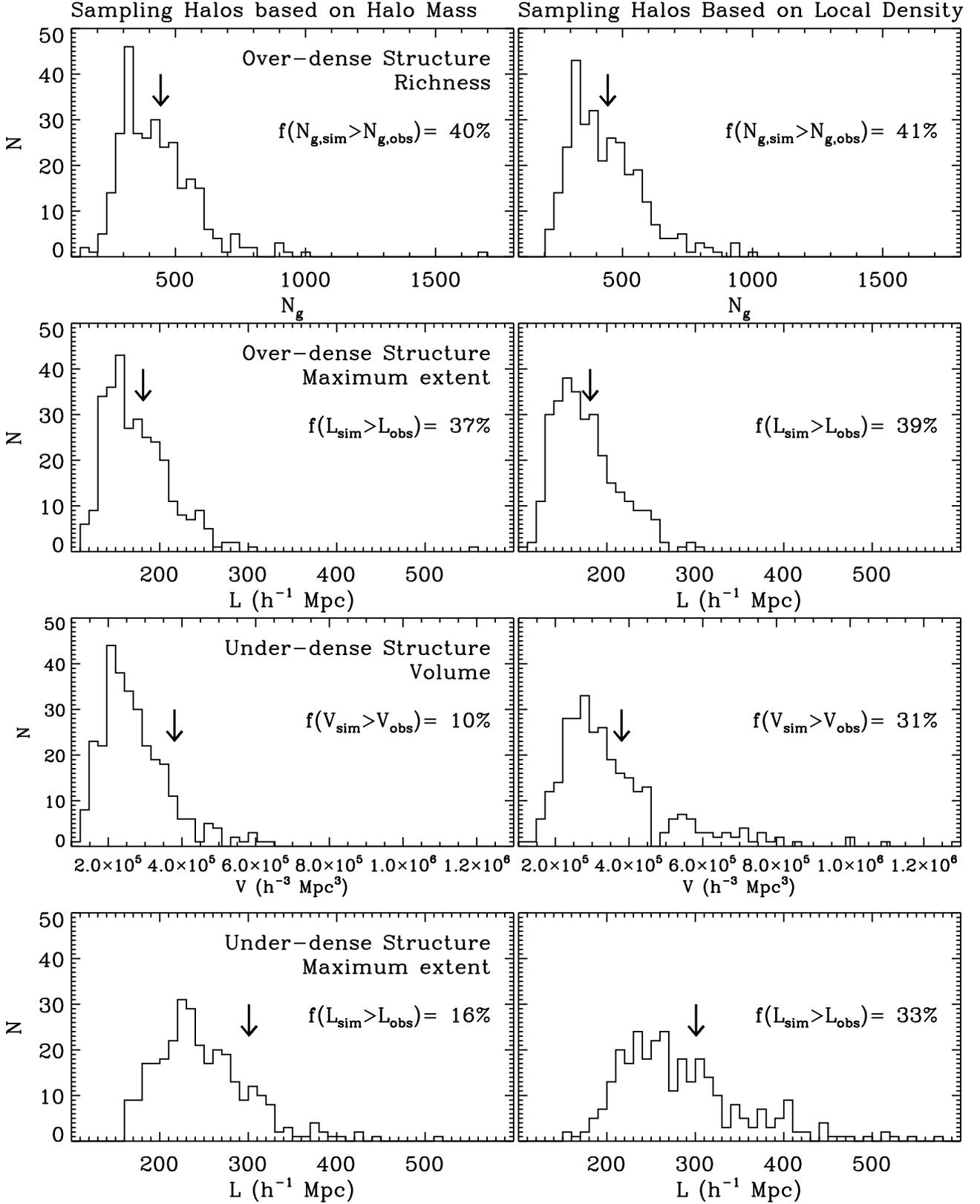}
\caption{({\it Left}) Distributions of the characteristics 
  of the largest structures
  derived from the 300 Horizon Run 4 mock surveys. 
The halo sampling is based on halo mass. 
The top two panels show the richness and size distributions
  of over-dense structures.
The bottom two panels show the 
  volume and size distributions of under-dense structures.
In each panel, the arrow indicates the parameter 
  for the largest structure in HectoMAP. 
The number in each panel indicates 
  the fraction of mock surveys that 
  contain a structure exceeding the scale for the observed structure. 
({\it Right}) Same as left panels, but for
  local density weighted sampling of simulated halos.
}\label{fig-largest}
\end{figure*}

Figure \ref{fig-largest} shows the distributions of the characteristics 
  of over-dense structures and under-dense structures (voids) 
  derived from the 300 Horizon Run 4 mock surveys. 
The left four panels show the results for sampling based on halo mass. 
The distributions are obviously not Gaussian; 
  they have long tails stretching toward large values of the parameter. 
They resemble log-normal distributions as expected on theoretical grounds 
  \citep{sd11}. 
In each panel the arrow indicates the parameter 
  for the largest structure in HectoMAP. 
We also indicate the fraction of mock surveys that 
  contain a structure exceeding the scale for the observed structure. 
It is striking that the void parameters lie 
  in the tails of the distribution
  whereas the dense structures lie relatively 
  near the median for the mock surveys.
This conclusion is consistent with the results 
  of statistical comparisons between observations and simulations 
  in Figures \ref{fig-den}-\ref{fig-voidsize};
  the only $\sim$2$\sigma$ rejection of similarity between the observations
  and simulations occurs for under-dense structures with sampling based on halo mass
  (see bottom left panels in Figures  \ref{fig-void} and  \ref{fig-voidsize}).

The right four panels of Figure \ref{fig-largest} show 
  the same distributions 
  but for local density weighted sampling of simulated halos. 
For the largest over-dense structure, 
  the relationship between the real universe and 
  the mock surveys is clearly insensitive to 
  the method of matching the mock surveys to the data. 
However, the lower two panels show that 
  a comparison based on local density weighted halo sampling
  has a substantial effect on the parameters 
  characterizing under-dense structures. 
Here the largest voids in HectoMAP are well within the range 
  of the distribution for the mock surveys. 
In the results based on a match of halo mass to stellar mass, 
  the parameters of the largest observed low density region 
  are in the upper 10$-$16\% of the simulated distribution: 
  when the selection of simulated halos is weighted by local density 
  to mimic the observational selection, 
  the characteristics of the largest low density region 
  are in the upper 31$-$33\% of the simulated distribution. 
This result is similar to the results of Section \ref{underlss}
  where the distribution of parameters 
  characterizing the low density regions is more similar 
  to local density weighted halo matching. 
For example, in Figure \ref{fig-void} (upper right) 
  the cumulative distribution of void volumes for the data 
  lies near the edge of the distribution for the mock surveys; 
  in Figure \ref{fig-void} (lower right), the HectoMAP distribution 
  is very close to the distribution for the mock surveys 
  analyzed with local density weighted halo sampling.

Use of the largest structures as a test of the models 
  is limited by the extended tails of the parameter distributions.
The largest dense structure in the mock surveys 
  (maximum extent: 560 $h^{-1}$ Mpc from mass weighted halo sampling) 
  is essentially as large as it can be to still fit within the HectoMAP region.
Although this structure is clearly an outlier 
  in the simulated distribution, 
  its mere presence is a warning about the robustness of 
  apparent inconsistencies based on observations of 
  a single large structure. 
With local density weighted halo sampling, 
  the maximum extent of the largest structure is much smaller, 
  310 $h^{-1}$ Mpc.
  
Use of the largest structure as a robust test of the models 
  would require observations of several well-separated (independent) volumes 
  comparable with HectoMAP. 
As emphasized in Sections \ref{overden} and \ref{underden}, 
  comparisons of different surveys of the real universe and 
  comparison between the models and the observations 
  are sensitive to the details of the analysis method; 
  the approach must be identical for clean comparisons.

\section{DISCUSSION}\label{discuss}

The combination of HectoMAP and Horizon Run 4 
  enables extension of statistical comparisons
  between the real and simulated universe to regions at $z>0.2$.
Surprisingly, the richness and size distributions
  of over-dense large-scale structures in HectoMAP 
  agree impressively with the Horizon Run 4 simulations.
The agreement also holds for under-dense structures or voids.
Thus the  standard $\Lambda$CDM model based on
  large-scale isotropy and homogeneity with 
  Gaussian primordial fluctuations
  successfully accounts for the observed large-scale structure
  in the galaxy distribution up to $z=0.44$. 

We note that the Horizon Run 4 simulations are dark matter only. 
Thus the remarkable agreement 
  between the observed and simulated universes 
  suggests that baryonic physics 
  does not play a major role on the scales we explore.
This conclusion is consistent with previous successful comparisons of 
  the observed large-scale galaxy distribution
  with halo distribution of dark matter only simulations.
  The SDSS/BOSS red galaxy samples, for example, imply a similar conclusion
  (e.g., \citealt{san12a,man13,nuza13,par14}).
The dense sampling  of  HectoMAP 
  and the huge volume covered by SDSS/BOSS 
  are complementary tests of the consistency between the models and the data.

Compared to previous studies, 
  there are several improvements in the comparison 
  of the data with the simulations here.
We examine the physical properties of
  both over- and under-dense large-scale structures 
  at intermediate redshift (i.e. $0.22<z<0.44$)
  by combining the dense, wide-field redshift survey data (i.e. HectoMAP)
  with similarly dense, large simulation data (i.e. Horizon Run 4).

We use  simulated true lightcones 
  drawn from the Horizon Run 4 simulation. 
Most other comparisons are based on snapshots from the simulations.
Because HectoMAP measures
  large-scale structure beyond the local universe,
  it is important to use true lightcones 
  analogous to the observed universe
  (see Section 3.1 in \citealt{kim15sim} for details on 
  the construction of lightcone and snapshots for Horizon Run 4).
  
The volume of Horizon Run 4 is about 3300 times larger than the volume
  of HectoMAP at $0<z<0.44$.
We thus can generate 300 non-overlapping HectoMAP-like mock surveys 
  from the Horizon Run 4 simulation.
These non-overlapping mock surveys from a huge simulation volume
  reinforce the statistical accuracy 
  in the comparison between observations and simulations. 
Because the match between the observations and the simulations is excellent, 
  the dispersion in the properties of the 300 simulations gives us 
  a robust measure of the probable error in the HectoMAP assessment 
  of the properties of large-scale structure. 
These 300 mock surveys enable exploration of 
  the largest over-dense structure and the largest void 
  for comparison with their HectoMAP counterparts. 
The observed structures are well within the range spanned by the mock surveys.

The physical properties of observed large-scale structures
  including richness, size and volume distributions
  strongly depend on the tracers and the method of identifying the tracers
  \citep{park12lss, park15lss}.
We thus emphasize the use of the same criteria
  in identifying large-scale structures in the observations and simulations 
  (i.e. friends-of-friends algorithm with the same linking length and
  the void-finding algorithm with the same ball radius)
  using comparable volume-limited samples of galaxies and halos
  matched according to number densities averaged over large volumes.
We also mimic the observational selection effects 
  (i.e. the choice of red galaxies) in the simulations
  by matching the local density surrounding halos 
  with those around the observed galaxies.
These controls enable a fair, reliable comparison
  of the physical properties of large-scale structures 
  in the real and simulated universe.

\section{CONCLUSIONS}\label{sum}

HectoMAP is a red-selected dense redshift survey,
  covering 53 deg$^{2}$ region of the sky 
  to a limiting magnitude of $r_{\rm Petro,0}=20.5$. 
Its high sampling density and large volume provides 
  a new basis for comparing large-scale structures 
  in the real and simulated universe in the redshift range $0.22 < z < 0.44$,
  covering a volume of $7.89\times10^6$ $h^{-3}$ Mpc$^3$.
We identify over- and under-dense large-scale structures in HectoMAP 
  and in 300 non-overlapping matched mock surveys drawn from 
  the Horizon Run 4 simulations. 
These observations and mock surveys provide 
  a well-controlled comparison of over-dense and under-dense
  large scale features of the galaxy distribution beyond the local universe.

We identify over-dense large-scale structure in HectoMAP 
  using the volume-limited sample 
  of red galaxies with $0.22<z<0.44$.
There are 87 over-dense structures with $N_{\rm member}\geq20$
  derived with a friends-of-friends algorithm and a  linking length of 
   $d_c=0.9d_{\rm mean}$ where $d_{\rm mean}$ is 9.01 $h^{-1}$ Mpc.
The richest and largest structure is at $z=0.36$ with
  a maximum extent of 181.1 $h^{-1}$ Mpc and 443 member galaxies.
  We construct 300 non-overlapping  mock surveys  from
  the Horizon Run 4 cosmological $N$-body simulation
  with the same halo number density 
  as in the HectoMAP volume-limited sample.
We identify over-dense large-scale structures in these mock surveys 
  using the same criteria as in HectoMAP.
The richness and size distributions of 
  the observed and simulated large-scale structures agree. 
The agreement is insensitive to  
  reasonable changes in the linking length 
  or in the method constructing volume-limited sample 
  of halos to mimic the data.

We also identify under-dense large-scale structures 
  using the same volume-limited
  samples of galaxies.
There are 72 under-dense structures with $N_{\rm pixel}\geq20$
  in the sample based on a ball radius of $r_{\rm ball}=1.64d_{\rm mean}$.
The largest structure is at $z=0.37$ with 
  a maximum extent of 300.8 $h^{-1}$ Mpc and
  a volume of $3.8\times10^5$ $h^{-3}$ Mpc$^3$.
We also identify under-dense structures in the 300 mock surveys 
  using the same criteria as in HectoMAP.
 The volume and size distributions of 
  observed and simulated voids are essentially 
  identical and insensitive to reasonable changes 
  in halo selection or in ball radius.

The size and richness (volume) distributions of observed
  over- and under-dense structures at $0.22<z<0.44$ in HectoMAP
  match those in the Horizon Run 4 dark matter only simulations.
The largest structures seen in HectoMAP are well 
  within the range of comparable structures
  identified in the 300 Horizon Run 4 mock surveys.
Thus on large scales, the features of the galaxy distribution 
  are insensitive to baryonic physics. 
The standard $\Lambda$CDM cosmological model
  accounts remarkably well for the structure we observe.
  
HectoMAP will eventually extend to 
   $r_{\rm Petro,0}<21.3$. The
  surface number density of galaxies will roughly double
  ($\sim1200$ per square degree).
We then expect to identify large-scale structures to $z\sim0.7$ 
  thus enabling exploration of the evolution of large-scale structures. 
To set the stage for the evolutionary considerations 
  it would be interesting to use the SDSS to construct local analogs 
  to the HectoMAP sample that could be analyzed 
  with essentially identical methods as a basis 
  for the assessment of evolution 
  from redshift zero to $z = 0.7$.

{\it Facility:} \facility{MMT Hectospec}

We thank the referee for a helpful and prompt report.
The Smithsonian Institution supports the research of M.J.G., D.G.F., M.J.K., P.B., M.C., S.T., and S.M..
H.J.Z. is supported by the Clay Postdoctoral Fellowship.
This work was supported by the Supercomputing Center/Korea Institute of Science and Technology Information with supercomputing resources including technical support (KSC-2013-G2-003). We thank Korea Institute for Advanced Study for providing computing resources (KIAS Center for Advanced Computation) for this work.
Funding for SDSS-III has been provided by the Alfred P. Sloan Foundation, the Participating Institutions, the National Science Foundation, and the U.S. Department of Energy Office of Science. The SDSS-III web site is http://www.sdss3.org/.
SDSS-III is managed by the Astrophysical Research Consortium for the Participating Institutions of the SDSS-III Collaboration including the University of Arizona, the Brazilian Participation Group, Brookhaven National Laboratory, Carnegie Mellon University, University of Florida, the French Participation Group, the German Participation Group, Harvard University, the Instituto de Astrofisica de Canarias, the Michigan State/Notre Dame/JINA Participation Group, Johns Hopkins University, Lawrence Berkeley National Laboratory, Max Planck Institute for Astrophysics, Max Planck Institute for Extraterrestrial Physics, New Mexico State University, New York University, Ohio State University, Pennsylvania State University, University of Portsmouth, Princeton University, the Spanish Participation Group, University of Tokyo, University of Utah, Vanderbilt University, University of Virginia, University of Washington, and Yale University.
This research has made use of the NASA/IPAC Extragalactic Database (NED) which is operated by the Jet Propulsion Laboratory, California Institute of Technology, under contract with the National Aeronautics and Space Administration.

\bibliographystyle{apj} 
\bibliography{ref_hshwang} 

\end{document}